\documentclass[a4paper,11pt, draftcls, onecolumn]{IEEEtran}
\usepackage{graphicx}
\usepackage{epstopdf}
\usepackage[utf8]{inputenc}
\usepackage[T1]{fontenc}
\usepackage[english]{babel}
\usepackage{amssymb}
\usepackage[T1]{fontenc}
\usepackage{hyperref}
\usepackage[english]{babel}

\usepackage{url}
\usepackage{multirow}
\usepackage{float}
\usepackage{cite}
\usepackage{amssymb,amsmath}
\usepackage[]{algorithm}
\usepackage[]{algorithmic}

\usepackage{setspace}

\usepackage{bm}
\usepackage{subcaption}
\usepackage{tikz}
\definecolor{gold}{rgb}{0.85,.66,0}

\begin{document}
\title{XL-MIMO  Energy-Efficient Antenna Selection under Non-Stationary Channels}
\author{Gabriel Avanzi Ubiali, Taufik Abr\~ao
\thanks{Copyright (c)2020. Personal use of this material is permitted.} 
\thanks{This work was supported in part by the Coordenac\~ao de Aperfeicoamento de Pessoal de N\'ivel Superior - Brazil (CAPES) - Finance Code 001, by the  Arrangement between the European Commission (ERC) and the Brazilian National Council of State Funding Agencies (CONFAP), CONFAP-ERC Agreement H2020, by the National Council for Scientific and Technological Development (CNPq) of Brazil under grant 310681/2019-7.}
}
	
	\maketitle
	\vspace{-1cm}
	
	
	\begin{abstract}
	Massive multiple-input-multiple-output (M-MIMO) is a key technology for 5G networks. Within this research area, new types of deployment are arising, such as the extremely-large regime (XL-MIMO), where the antenna array at the base station (BS) {has extreme dimensions}. As a consequence, spatial non-stationary properties appear {as} the users see only a portion of the antenna array, which is called visibility region {(VR)}. In this {challenging transmission-reception} scenario, an algorithm to select the {appropriate  antenna-elements} for processing the received signal of a given user {in the uplink (UL)}, as well as to transmit the signal of this user during downlink {(DL)} is proposed. The advantage of not using all the available {antenna-elements} at the BS is the {computational burden and circuit power consumption reduction}, improving the energy efficiency {(EE) substantially}. Numerical results {demonstrate} that one can increase the {EE} without compromising considerably the spectral efficiency {(SE). Under few active users scenario, the performance of the XL-MIMO system shows that the EE} is maximized using less than 20\% of the antenna-elements of the array, without compromising the SE severely.
	\end{abstract}
	

	\section{Introduction}\label{sec:intro}
	{Massive multiple-input-multiple-output (M-MIMO) is one of the key technologies for 5G networks}\cite{EMIL2014_5directions_for_5G}, which permits that more than one single user transmits simultaneously with high spectral and energy efficiencies and using the same spectrum, {\it i.e.}, many antennas simultaneously serve many users using the same time-frequency resource\cite{Marzetta2010}. {In MIMO networks, the base station (BS) estimates the channel coefficients and employs a transmit precoding scheme in the downlink (DL) and a receive combining scheme in the uplink (UL), giving each user a different spatial signature} \cite{marzetta2016fundamentals,chockalingam}. M-MIMO wireless communication is a special case of MIMO systems using hundreds of antennas at the BS, providing sufficient spatial dimensions to uncover the fundamental properties of M-MIMO: channel hardening, large array gain and asymptotic inter-terminal channel orthogonality {(favorable propagation)} \cite{carvalho2019nonstationarities}. {Thus, it can provide large improvements over traditional systems in both energy and spectral efficiencies.}
	
As the number of BS antennas increases, it is possible to focus the transmission and reception of signal energy into ever-smaller regions of space, which brings huge improvements in throughput and EE. {However, it may come with the computational complexity increasing, as well as with the increasing of implementation cost and power consumption, what advocates for new deployments that take real advantage from increasing the number of BS antennas to the order of hundreds or thousands, without severe problems due to the holdbacks above cited. Moreover, in order to make real advantage from the deployment of such a large number of antenna elements, it is desirable to distribute them over a substantially large area in order to increase the antenna separation and coverage \cite{massivemimobook}. One potential approach is the extremely-large MIMO (XL-MIMO) regime, where the antenna array is integrated into large building structures\cite{carvalho2019nonstationarities}.}
	
{When a moderate number of (several tens of) antennas is compactly deployed in the BS, the entire array will receive approximately the same amount of energy from each user, \textit{i.e.}, the channel is spatially stationary\cite{ExpPropDetector_XL-MIMO}. On the other hand, in the XL-MIMO regime, different parts of the array may observe the same propagation paths with varying power and phases or distinct propagation paths. Then, the majority of energy received from a specific user concentrates on small portions of the entire array, which is a channel property called spatial \textit{non-stationarity}, which has been observed by recent channel measurements \cite{SparseChannelEstimation,non_stat_measurements1,non_stat_measurements2} and can be introduced in the channel model by using the concept of visibility region (VR) \cite{XL_MIMO_letter, carvalho2019nonstationarities, amiri2018extremely,2100channelmodel}. However, it is worth saying that the density of VRs influences on the size of the portion of the array that the user can see. For instance, if the VR density is sufficiently high, all portions of the array are able to receive some signal energy.}
	
{The conventional M-MIMO signal processing architecture is centralized at the BS, what means that the signals are received at the BS (UL) and transmitted by the BS (DL) deploying all elements of antennas. Then, the associated computational complexity becomes a challenge when employing extremely large arrays, specially in crowded scenarios, due to the need to transfer excessively large amounts of data received by the array to the processing unit \cite{ExpPropDetector_XL-MIMO}. A promising solution is to use only a portion of the whole element-antennas array to perform receive combining as well as the transmit precoding to each user.}

{Hence, by appropriately selecting a subset of BS antennas to communicate with each user, the receiver is able to capture almost the totality of the energy transmitted by that user, while reducing the interference coming from the other users and therefore potentially increasing the spectral efficiency (SE). Furthermore, by having higher SE and reduced power consumption, one can obtain higher EE. It is important to highlight that there is a growing concern about how to improve the EE in wireless communications, as the increasing data rates and the increasing number of users connected to the network increase substantially the overall energy consumption\cite{Huang.2018}. For this reason, in this work we take both SE and EE as performance metrics to be analyzed in section \ref{sec:results}. Finally, on can summarize the benefits of the VR-based subarray antenna selection architecture as: a) computational complexity reduction; b) overall energy saving by activating a reduced number of antenna-elements;  c) as a result an EE increasing; d) while potentially improving the overall system SE by selecting appropriate antenna-elements associated to the each user VR.}

{Considering that the antenna array experiences spatial non-stationarities when using large-aperture arrays, in \cite{near_field}, authors propose two new channel estimation methods that, besides estimating the channel vector, obtains the position of the scatterers and the visibility regions, which may be useful for transceiver design.} In \cite{XL_MIMO_letter}, authors study the impact of spatial non-stationarity where the channel energy is concentrated on a portion of the array, \textit{i.e.}, the VR, in terms of signal-to-interference-plus-noise ratio (SINR) performance. In \cite{carvalho2019nonstationarities}, authors show that, when M-MIMO systems operate in extra-large scale regime, several important MIMO design aspects change, due to spatial non-stationarities, where the users see only a portion of the array and, inside the VR, different parts of the array see different propagation paths. Moreover, three low-complexity data detection algorithms are proposed in \cite{amiri2018extremely} as candidates for uplink communication in XL-MIMO systems.

{In \cite{ExpPropDetector_XL-MIMO}, authors designed an efficient detector for extra-large-scale massive MIMO systems with the subarray-based processing architecture, by extending the application of the expectation propagation principle. Their analysis is based on bit error rate (BER) performance. A different approach to the antenna selection methodology  proposed herein, in \cite{amiri2020deep}, authors propose a design based on machine learning to select a small portion of the array that contains the beamforming energy to the user, aiming to overcome the prohibitive complexity of XL-MIMO systems. They provide numerical results in terms of sum-rate performance. To the best of author's knowledge, this is the first work addressing the subarray-based processing architecture, which remarkably provides huge reduction in the computational complexity, through the point of view of improving the overall system EE.}

\noindent{\textbf{\textit{Contribution}}. We deal with an XL-MIMO system eqquiped with a subarray-based processing architecture, in order to reduce the overall system computational complexity. In such scenario, we propose a novel algorithm to judiciously select the antenna-elements subarray that will communicate with each user, aiming at obtaining higher EE while reducing the power consumption, when compared to the whole antenna array activation to communicate with every user. Thus, the {\it contribution} of this work can be summarized as: \textbf{\textit{i}}) we propose an antenna selection procedure to improve simultaneously the overall system EE by reducing the power consumption, taking into account the spatial non-stationarity assumption while taking advantage of the VRs features; \textbf{\textit{ii}}) the proposed algorithm also provides a considerable computational complexity reduction;
\textbf{\textit{iii}}) a comprehensive analysis development on how the proposed procedure impacts the system performance is developed, highlighting and characterizing its benefits when comparing to the condition of using the entire antenna array to communicate with every user.}

{The remainder of the paper  is organized as follows.  The adopted XL-MIMO channel model and the channel estimation procedure is developed in Section \ref{sec:system_model}; this section also provides the ergodic UL and DL spectral efficiencies expressions, based on the signal-to-interference-plus-noise ratio (SINR). Section \ref{sec:processing_algorithm} focuses on the proposed antenna selection procedure, as well as on the computational complexity aspects. The EE definition and a detailed circuit power model are discussed in Section \ref{sec:SE&EE}. Section \ref{sec:results} examines  numerical results corroborating our findings, while the main conclusions are presented in Section \ref{sec:conclusion}.}

\section{System Model}\label{sec:system_model}

    We consider the UL and the DL of a single-cell multiuser XL-MIMO system with an $M$-antenna BS and $K$ single-antenna users at each cell, operating over a bandwidth of $B$ Hz. The channel estimates are acquired via UL synchronous pilot transmission. The time-division duplex (TDD) operation mode was chosen because of its advantages over the frequency-division duplex (FDD) mode. TDD does not require quantized channel state information (CSI) to be sent by the BS to the user via feedback, because of channel reciprocity, avoiding excessive overhead \cite{Flordelis_2018, marzetta2016fundamentals}.
	
\begin{figure}[!htbp]
    \centering
    \includegraphics[width=.6\textwidth]{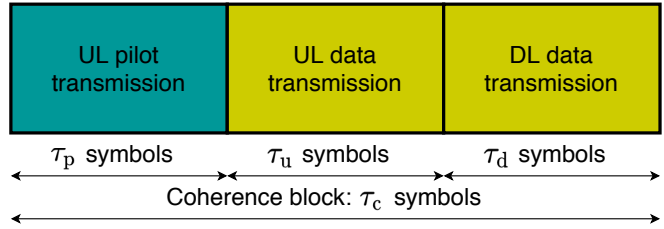}
    \caption{{Each coherence block is divided into the UL pilot transmission for channel estimation purpose, the UL data transmission, and the DL data transmission.}}
    \label{CoherenceBlock}
    \end{figure}

	The channel coherence time ($T_\text{C}$) is divided into UL pilot, UL data and DL data transmission, as Figure \ref{CoherenceBlock} shows. The number of symbols that fits in a channel coherence block is $\tau_\text{c} = T_\text{C} B_\text{C}$, being $B_\text{C}$ the coherence bandwidth \cite{Marzetta2010}. In order to estimate the channel, each of the $K$ users of a given cell is assigned a different pilot sequence. There are $\tau_\text{c}$ symbols per coherence block, of which $\tau_\text{p}$ are dedicated to UL pilot transmission, $\tau_\text{u} = \epsilon_\text{u}\left(\tau_\text{c}-\tau_\text{p}\right)$ symbols are dedicated to UL data transmission and $\tau_\text{d} = \epsilon_\text{d}\left(\tau_\text{c}-\tau_\text{p}\right)$ symbols are dedicated to DL data transmission, where $\epsilon_\text{u} + \epsilon_\text{d} = 1$. The number of available orthogonal pilot sequences is equal to its length ($\tau_\text{p}$). As we need $K$ sequences, we can take $\tau_\text{p} = K$. Thus, the time required for pilots is proportional to the number of users served. The number of users that can be served is therefore limited by the coherence time, which itself depends on the mobility of the users \cite{Marzetta2010}.
    
\begin{figure}[!htbp]
    \centering
    \includegraphics[width=.75\textwidth]{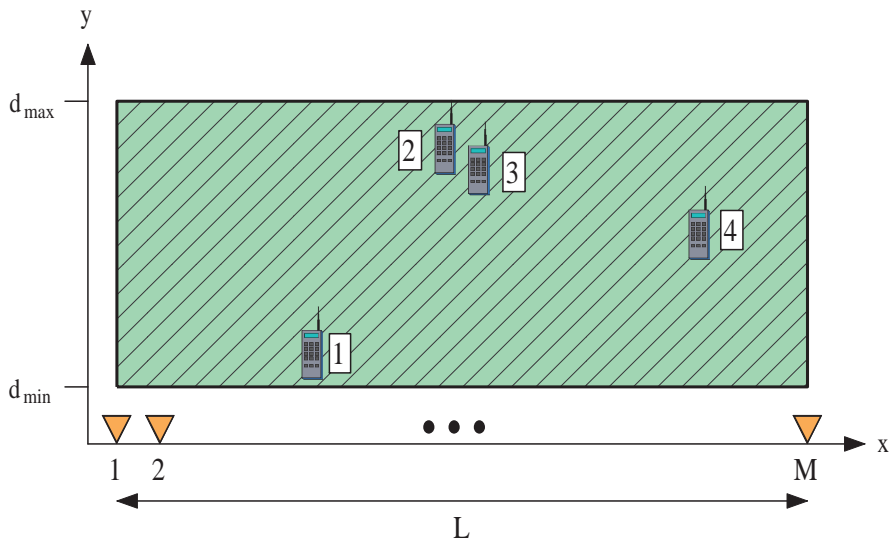}
    \caption{Typical system spatial configuration.} 
    \label{fig:scenario}
    \end{figure}

{The antenna-elements are uniformly spaced over a uniform linear $L$-length array containing $M$ elements ($M$-ULA). The coordinates of the first and the $M$-th antennas are $(0,0)$ and $(L,0)$, respectively, which means that, when $M\geq1$, the spacing between the antennas is $\frac{L}{M-1}$. The users are placed over a rectangle that extends along the antenna array in one dimension and between a minimum ($d_\text{min}$) and a maximum distance ($d_\text{max}$) in the other dimension, following a uniform distribution over this area. The coordinates of the $m$-th antenna are denoted by $(a_m,0)$, and its distance to the $k$-th user is denoted by $d_{mk}$. A typical system configuration is represented by Figure \ref{fig:scenario}.}
    
    \subsection{XL-MIMO Channel Model}
    \label{subsec:channel}
    {In XL-MIMO systems, spatial non-stationarities occur frequently since typically $d_{\max} <L$, what means that a given user probably sees only a small part of the antenna array. The propagation environment contains numerous objects reflecting the signal, which are called scattering points. Each scattering point has an associated VR. Herein, we assume that each user can see the antenna array through $N_\text{c}$ different VRs. The $i$-th VR that is visible to the $k$-th user extends from the $c_{ik}$-th to the $(c_{ik} + N_{ik})$-th antenna and is denoted by $\mathcal{C}_{ik} = \{c_{ik}\ ,\ c_{ik} + 1\ ,\ \dots\ ,\ c_{ik} + N_{ik}\}$. These VRs may overlap, and the set $\mathcal{C}_k = \mathcal{C}_{1,k}\ \cup\ \dots\ \cup\ \mathcal{C}_{N_\text{C},k}$ contains the indices of the antennas that are visible to the $k$-th user. Considering that each user sees the antenna array through more than one VR, it is a way of taking into account that not only one subset of the array containing contiguous antenna elements may be visible. Indeed, the mobile user may see more than one portion of the array. For instance, if $M = 100$ in  Figure \ref{fig:scenario}, it may see antennas 10 to 20 and 25 to 40, which means there is something in the propagation environment preventing the signal from reaching antennas 21 to 24.}
    
    {In general, the distances between a given user and all the BS antennas are considered to be the same. However, in the XL-MIMO scenario, as the length of the antenna array is not negligible, the pathloss varies throughout the array, mainly when $d_{\max} \ll L$. The pathloss coefficient between the $m$-th antenna and the $k$-th user is given by:}
\begin{equation}
{b_{mk} = \frac{b_0}{(d_{mk})^{\gamma}}}  
\label{b}
\end{equation}
    {where $\gamma \geq 2$ is the pathloss exponent and $b_0$ determines the median channel gain at a reference distance of 1 m\cite{massivemimobook}. The parameters $b_0$ and $\gamma$ are functions of the carrier frequency, antenna gains, and vertical heights of the antennas, which are derived from fitting \eqref{b} to measurements \cite{massivemimobook}.}
{Finally, the channel vector $\textbf{h}_k = [h_{1k}~\dots~h_{Mk}]^\text{T}$ between the BS and the $k$-th user is given by:}
	\begin{equation}
	{\textbf{h}_k = \textbf{a}_k\odot\sqrt{\textbf{b}_k}\odot\overline{\textbf{h}}_k}
	\label{h_k}
	\end{equation}
    {where $\textbf{b}_k$ is a vector whose $m$-th element is $b_{mk}$, and $\overline{\textbf{h}}_k$ is the independent Rayleigh fading component, which accounts for the short-scale fading and follows a complex-Gaussian  distribution $\overline{\textbf{h}}_k \sim \mathcal{CN}(\textbf{0}_M, \textbf{I}_M)$. The coefficient $a_{mk} = [\textbf{a}_k]_m$ indicates whether the $m$-th antenna is visible to the $k$-th user ($a_{mk}$ = 1) or not ($a_{mk}$ = 0), and is given by:}
	\begin{align}
	{a_{mk} = 
	\begin{cases}
	1 \quad & m \in \mathcal{C}_k \\
	0 & \text{otherwise}.
	\end{cases}}
	\label{alpha_non_statio}
	\end{align}
	%

\subsection{UL Pilot Transmission} \label{subsec:system_model_ULpilot}
We assume the channel is estimated via UL synchronous pilot transmission, which means that all users simultaneously send pilot sequences from the same pilot codebook. They have length $\tau_\text{p}$ and form an orthogonal set. Herein, it is assumed that each user is assigned a different pilot sequence. Then, the pilot sequences' set is $\bm{\Psi} = [\bm{\psi}_1 \ldots \bm{\psi}_K] \in \mathbb{C}^{\tau_\text{p} \times K}$ and the orthogonality condition states that $\bm{\Psi}^\text{H}\bm{\Psi} = \tau_\text{p}\textbf{I}_{\tau_\text{p}}$, {\it i.e.}:
    \begin{equation}
    \bm{\psi}_i^\text{H}\bm{\psi}_k = 
    \begin{cases}
    \tau_\text{p} \quad & i = k, \\
    0 \quad & i \neq k.
    \end{cases}
    \end{equation}

    During the UL pilot transmission, the $k$-th user transmits the pilot sequence $\bm{\psi}_k \in \mathbb{C}^{\tau_\text{p}}$, with transmit power $p_\text{p}$. The elements of $\bm{\psi}_k$ are scaled by $\sqrt{p_\text{p}}$, forming the signal $\textbf{s}_k = \sqrt{p_\text{p}}\bm{\psi}_k^\text{H}$, to be transmitted over $\tau_\text{p}$ UL samples. As a result, the BS receives the signal $\textbf{Y}^\text{p} \in \mathbb{C}^{M\times \tau_\text{p}}$:
	\begin{equation}
	\textbf{Y}^\text{p} = \sum_{i=1}^K \sqrt{p_\text{p}} \textbf{h}_i \bm{\psi}_i^\text{H} + \textbf{N}^\text{p}
	\end{equation}
	where $\textbf{N}^\text{p} \in \mathbb{C}^{M\times\tau_\text{p}}$ is the noise matrix at the receiver of the BS with i.i.d. entries following a complex normal distribution with zero mean and variance $\sigma_\text{UL}^2$.
	
{As the information about which antennas are visible for each user is unknown, it might be necessary to consider obtaining the channel estimates by using estimators that require no prior statistical information, such as the least-squares (LS). Moreover, as we consider that each user is assigned a different pilot sequence and these sequences are mutually orthogonal, there is no pilot contamination. Hence, the imperfections on the channel estimates are just due to the noise power at the BS antennas during the UL pilot transmission. It advocates for using LS channel estimation rather than MMSE. The LS estimate of $\textbf{h}_k$ is attained by\cite{massivemimobook}:}
	\begin{subequations}
		\begin{align}
		\hat{\textbf{h}}_k
		&= \frac{1}{\tau_\text{p}\sqrt{p_\text{p}}}\textbf{Y}^\text{p}\bm{\psi}_k \label{h_est}\\
		&= \textbf{h}_k + \frac{1}{\tau_\text{p}\sqrt{p_\text{p}}}\textbf{N}^\text{p}\bm{\psi}_k
		\label{h_est2}
		\end{align}
		\label{hest}
	\end{subequations}
	\indent The last term in \eqref{h_est2} is the equivalent noise vector, which adds imperfections to the channel estimates and follows a complex normal distribution: $\textbf{N}^\text{p}\bm{\psi}_k \sim \mathcal{CN}(\textbf{0}_M, \tau_\text{p}\sigma_\text{UL}^2\textbf{I}_M)$. Finally, the estimated channel matrix is $\hat{\textbf{H}} = [\hat{\textbf{h}}_1~\dots~\hat{\textbf{h}}_K]$, while $\textbf{H} = [\textbf{h}_1~\cdots~\textbf{h}_K] \in \mathbb{C}^{M\times K}$ is the true channel matrix. {According to equation \eqref{h_est}, the channel estimation process corresponds to the inner product of $M$ complex vectors of length $\tau_\text{p}$, requiring $M\tau_\text{p}$ multiplications between complex numbers (or $3M\tau_\text{p}$ multiplications between real numbers\footnote{Consider $x = a + jb$ and $y = c + jd$. The hardware implementation of the complex multiplication $xy = ac - bd + j[(a + b)(c + d) - ac - bd]$ involves 3 real multiplications and 5 real sums. Only those will be considered, due to their very greater hardware complexity compared to the real sum operation.}) to estimate the channel vector of each of the $K$ users. Herein, we consider that both multiplication and division between real numbers correspond to 1 floating-point operation (flop). As the channel estimation process is performed once per coherence block, its computational complexity, defined in number of flops per coherence block [fpcb], is:}
\begin{equation}
{C_\text{CE} = 3MK\tau_\text{p} \qquad\qquad \text{[fpcb]}}
\label{C_CE}
\end{equation}

\subsection{UL Data Transmission}\label{subsec:system_model_ULdata}
{The received signal $\textbf{r} \in \mathbb{C}^M$ at the BS during the UL data transmission is:}
\begin{align}
{\textbf{r} = \sum_{k=1}^K \textbf{h}_k x_k + \textbf{n}}
\label{rUL}
\end{align}
{where $x_k$ is the signal sent by the $k$-th user and $\textbf{n} \sim \mathcal{CN}( \mathbf{0}_M, \sigma_\text{UL}^2 \textbf{I}_M)$ contains the noise received at the BS antennas. By utilizing a suitable combining vector, $\textbf{v}_k \in \mathbb{C}^M$, the BS detects the $k$-th user's signal as follows:}
\begin{align}
{y_k	= \textbf{v}_k^\text{H}\textbf{r}}
	\label{receiveSP}
	\end{align}
	
{Herein we consider the two simplest types of linear processing for receive combining: zero-forcing (ZF) and maximum-ratio (MR), which are respectively defined by:}
\begin{align}
	{\textbf{V} = \hat{\textbf{H}} ( \hat{\textbf{H}}^\text{H} \hat{\textbf{H}})^{-1}}
	\label{ZF}
	\end{align}
    {and}
	\begin{equation}
{\textbf{V} = \hat{\textbf{H}}}
	\label{MR}
	\end{equation}
{where the matrix $\textbf{V} = [\textbf{v}_1\dots\textbf{v}_K]$ is the collection of the combining vectors. ZF induces considerably smaller intra-cell interference than MR, yielding significantly better performance under interference-limited conditions, which is normally the case. On the other hand, it increases computational complexity significantly when employing large antenna arrays, due to multiplications of complex numbers, and when serving a great number of users, due to the size of the $K\times K$ matrix that is inverted\cite{marzetta2016fundamentals,massivemimobook}. Due to these particularities, 
it is interesting to compare the system performance in terms of SE and EE considering such linear processing techniques.}

The channels are practically constant within a coherence block, while the signals and noise take new realization at every sample. Then, the instantaneous SINR is actually an expectation over one coherence block, what means that $p_k^\text{UL} = \mathbb{E}\{|x_k|^2\}$ - which is the UL transmit power of the $k$th user - and $\sigma_\text{UL}^2$ will be taken instead of the instantaneous values of $|x_k|^2$ and $|n_k|^2$, respectively. Thus, one can define the SINR of the $k$th user during the UL data transmission as
	\begin{equation}
	\gamma_k^\text{UL} = \frac{p_k^\text{UL}|\textbf{v}_k^\text{H}\textbf{h}_k|^2}{\sum\limits_{\substack{i=1\\i\neq k}}^K p_i^\text{UL}|\textbf{v}_k^\text{H}\textbf{h}_i|^2 + \sigma_\text{UL}^2||\textbf{v}_k||^2}
	\label{SINR_k_UL}
	\end{equation}
	
    As a result, the UL ergodic spectral efficiency is defined by\cite{massivemimobook}:
    \begin{equation}
    \text{SE}_\text{UL} = \frac{\tau_\text{u}}{\tau_\text{c}}\sum_{k=1}^K\mathbb{E}\{\text{log}_2 (1 + \gamma_k^\text{UL})\}
    \label{SE_UL}
    \end{equation}

    \subsection{DL Data Transmission}
    \label{subsec:system_model_DLdata}

{In the DL, the information to be transmitted by the BS to the $k$-th user, $x_k$, needs to be precoded, by using the precoding vector $\textbf{w}_k$. The signal to be transmitted, denoted by $\textbf{s} \in \mathbb{C}^M$, is generated as:}
\begin{align}
{\textbf{s} = \sum_{k=1}^K \textbf{w}_k x_k}
\label{transmitSP}
\end{align}

{We denote the matrix containing the collection of the precoding vectors by $\textbf{W} = [\textbf{w}_1~\cdots~\textbf{w}_K] \in \mathbb{C}^{M\times K}$. The UL-DL duality motivates a simple precoding design principle: selecting the DL precoding vectors as the normalized version of their respective combining vectors\cite{massivemimobook}:}
\begin{equation}
{\textbf{w}_k = \frac{\textbf{v}_k^*}{||\textbf{v}_k||}}
	\label{w_k}
	\end{equation}

	Assuming no receive combining, user $k$ receives the signal:
	\begin{align}
	y_k = \textbf{h}_k^\text{T}\textbf{s} + n_k
	\end{align}
	where the received noise follows the distribution $n_k \sim \mathcal{CN}(0, \sigma_\text{DL}^2)$. Analogously to \eqref{SINR_k_UL} for the UL, the SINR of the $k$th user in the DL data transmission is defined as:
	\begin{equation}
	\gamma_k^\text{DL} = \frac{p_k^\text{DL}|\textbf{h}_k^\text{T}\textbf{w}_k|^2}{\sum\limits_{\substack{i=1\\i\neq k}}^K p_i^\text{DL}|\textbf{h}_k^\text{T}\textbf{w}_i|^2 + \sigma_\text{DL}^2}
	\label{SINR_k_DL}
	\end{equation}
	where $p_k^\text{DL}$ is the downlink transmit power assigned for user $k$, i.e., $\sqrt{p_k^\text{DL}}$ scales the vector $\textbf{w}_k$, which has unit norm. Finally, the DL ergodic spectral efficiency is given by\cite{massivemimobook}:
	\begin{equation}
	\text{SE}_\text{DL} = \frac{\tau_\text{d}}{\tau_\text{c}}\sum_{k=1}^K\mathbb{E}\{\text{log}_2 (1 + \gamma_k^\text{DL})\}
	\label{SE_DL}
	\end{equation}

	\section{Antenna Selection for  Combining and Precoding in XL-MIMO}\label{sec:processing_algorithm}
	{In this section, we propose an algorithm to select the antenna-elements in an XL-MIMO system for received signal processing (combiner) of a given user during the UL and transmit the signal (precoder) of this user during the DL. One advantage of not using all the $M$ available antennas is the reduction of the computational complexity and the circuit power consumption (Tx and Rx operations), as less antennas are active at the same time. Furthermore, the throughput potentially increases, because the interference power at the receivers decreases since each antenna individually does not serve all the $K$ users simultaneously.}
	
	\subsection{{HRNP-based Antenna Selection Criterion and Algorithm}}
	\label{subsec:alg}
	
	{First, Algorithm \ref{alg} computes the vector $\boldsymbol{\theta}_k \in \mathbb{C}^M$, which is a quantitative indicator of the quality of the channel between the $k$-th user and each of the $M$ antennas as:}
	\begin{equation}
	{\theta_{mk} = \frac{|\hat{h}_{mk}|^2}{\sum\limits_{\substack{i=1\\i\neq k}}^K|\hat{h}_{mi}|^2},\qquad\qquad  m= 1,2,\ldots, M}
	\label{theta}
	\end{equation}
	{where $\hat{h}_{mk} = [\hat{\textbf{h}}_k]_m$ and $\theta_{mk} = [\boldsymbol{\theta}_k]_m$. A high signal intensity may be obtained when $|\hat{h}_{mk}|^2$ is strong. Selecting the $N$ strongest $\theta_{mk}$ values among $m=1, \ldots, M$ in \eqref{theta} for each user, provides the {\it highest received normalized power} (HRNP)  antenna selection criterion. On the other hand, the terms $|\hat{h}_{mi}|^2$, $i\neq k$, are related to the interference intensity. Higher $\theta_{mk}$ values are therefore associated to higher SINRs on the signal detection, as defined by the equations \eqref{SINR_k_UL} and \eqref{SINR_k_DL}, and consequently higher SE and EE.}
	
\noindent{Second, the Algorithm \ref{alg} obtains the set $\mathcal{D}_k$ (lines 5--10), which contains the indices of the $N$ antennas with the highest $\theta_{mk}$ values. Only these $N$ antenna-elements are activated for user $k$. Lastly, Algorithm \ref{alg} computes the receive combining and the transmit precoding vectors of the $k$-th user (lines 11--14) based on the matrix $\hat{\textbf{H}}_k\in\mathbb{C}^{N\times K}$, which contains all the columns of the estimated channel $\hat{\textbf{H}}$ but only the rows corresponding to the elements of the set $\mathcal{D}_k$.}
	
As the set $\mathcal{D}_k$ contains the indices of the antennas that are active for the $k$th user, the superset $\mathcal{D} = \mathcal{D}_1 \cup \dots \cup \mathcal{D}_K$ contains all the indices of the antennas that are active for any user. {The number of elements in $\mathcal{D}$, denoted by $N_\text{act}$, corresponds to the total number of active elements of antenna.} Notice that the rows of the combining and the precoding matrices corresponding to the antennas whose indices are not in the set $\mathcal{D}_k$ are set equal zero.

\begin{algorithm}[hbt]
\caption{Antenna selection (AS) for receive combining and transmit precoding}
\label{alg}
\hspace{\algorithmicindent}~~{\textbf{Input:} $M$, $N$, $K$, $\hat{\textbf{H}}$}\\
\hspace*{\algorithmicindent}~~{\textbf{Output:}  $\textbf{V}$, $\textbf{W}$}
\begin{algorithmic}[1]
	\STATE {Initialize the combining matrix $\textbf{V}$ with $\textbf{0}_{M\times K}$}
\FOR{$k$ = 1 to $K$}
\STATE {Compute vector $\boldsymbol{\theta}_k$ via eq. \eqref{theta}}
			\STATE Reinitialize the set of the indices of the antennas: $\mathcal{M} = \{1,\dots,M\}$
			\STATE Initialize $\mathcal{D}_k = \emptyset$
			\FOR{$n$ = 1 to $N$}
			\STATE find $m^* = \text{arg} \underset{m\in \mathcal{M}}{\text{max}}\ {\theta_{mk}}$
			\STATE $\mathcal{M} = \mathcal{M} \backslash m^*$
			\STATE $\mathcal{D}_k = \mathcal{D}_k \cup \{m^*\}$
			\ENDFOR
			\STATE $\hat{\textbf{H}}_k = \hat{\textbf{H}}(\mathcal{D}_k,:)$
			\STATE {If MR is selected: $\textbf{V}_\text{MR}(\mathcal{D}_k,k) = \hat{\textbf{H}}_k(:,k)$}
			\STATE {If ZF is selected: $\textbf{V}_\text{ZF}(\mathcal{D}_k,k) = [\hat{\textbf{H}}_k(\hat{\textbf{H}}_k^\text{H}\hat{\textbf{H}}_k)^{-1}]_{(:,k)}$}
			\STATE {$\textbf{W}(:,k) = \frac{\textbf{V}(:,k)^*}{||\textbf{V}(:,k)||}$}
		\ENDFOR
	\end{algorithmic}
	\end{algorithm}

\subsection{{Computational Complexity}} \label{subsec:comp}
	
{We first address the complexity of computing the ZF combining vectors (line 13, Algorithm \ref{alg}). Recalling that $\hat{\textbf{H}}_k\in\mathbb{C}^{N\times K}$, the multiplication of $\hat{\textbf{H}}_k^\text{H}$ by $\hat{\textbf{H}}_k$ requires $\frac{K^2+K}{2}N$ complex multiplications\footnote{{Being $\textbf{A}\in\mathbb{C}^{a\times b}$ and $\textbf{B}\in\mathbb{C}^{b\times c}$, the multiplication $\textbf{A}\textbf{B}$ requires $ac$ inner products between $b$-length vector, what corresponds to $abc$ complex multiplications. However, if $\textbf{B} = \textbf{A}^\text{H}$, the Hermitian symmetry is utilized. Thus, only the $a$ diagonal elements of $\textbf{A}\cdot\textbf{B}$ and half of the $a^2-a$ off-diagonal elements need to be computed, what gives $\frac{a^2+a}{2}b$ complex multiplications\cite{massivemimobook}.}}, using the Hermitian symmetry. When the inverse of a matrix is multiplied by another matrix, the $\textbf{LDL}^\text{H}$ decomposition can be used to achieve an efficient hardware implementation \cite{massivemimobook}. The decomposition of $\hat{\textbf{H}}_k^\text{H}\hat{\textbf{H}}_k$ requires $\frac{K^3-K}{3}$ complex multiplications\cite{massivemimobook}. Finally, we need to multiply the matrix $\hat{\textbf{H}}_k$ by the $k$-th column of the matrix $(\hat{\textbf{H}}_k^\text{H}\hat{\textbf{H}}_k)^{-1}$, which requires $KN$ complex multiplications plus $K$ complex divisions to compute $\textbf{D}^{-1}$\cite{massivemimobook,boyd}. Considering complex multiplications and complex divisions to correspond to 3 and 7 flops\footnote{{Considering $x = a + jb$ and $y = c + jd$, then $\frac{x}{y} = \frac{xy^*}{yy^*} = \frac{xy^*}{|y|^2}$, while the computation of $xy^*$ requires 3 real multiplications. The computation of $|y|^2 = c^2 + d^2$ requires 2 real multiplications. Finally, the complex division $\frac{xy^*}{|y|^2}$ corresponds to 2 real divisions, making a total of 7 real operations.}}, respectively, the computation of the combining vector $\textbf{v}_k$ has a complexity of $3\left(\frac{K^2+K}{2}N + \frac{K^3-K}{3} + KN\right) + 7K$ flops per coherence block. Thus, the computational complexity to obtain the whole combining matrix $\textbf{V}$ is given by:}
\begin{equation}
{C_\text{SP-C}^\text{UL-ZF} = K^4 + \frac{3}{2}K^3N + \frac{9}{2}K^2N + 6K^2}   \qquad {\text{[fpcb]}}
\label{C-SP-UL-ZF}
\end{equation}

{As defined in \eqref{MR}, MR combining does not require multiplications or divisions, because it is given directly from the channel estimates (from Algorithm \ref{alg}, one can see that the $k$-th column of the MR combining matrix is simply a copy of the $k$-th column of $\hat{\textbf{H}}_k$). However, in practical implementations, we typically normalize the combining vector such that $\textbf{v}_k^\text{H}\textbf{h}_k$ in front of the desired signal $x_k$ is close to one. Thus, this normalization requires 1 complex division per user\cite{massivemimobook}, resulting in a total of $7K$ flops per coherence block. Finally, the complexity of computing the MR combining matrix is given by:}
\begin{equation}
{C_\text{SP-C}^\text{UL-MR} = 7K} \qquad\qquad {\text{[fpcb]}}
	  \label{C-SP-UL-MR}
	\end{equation}
	
{The precoding vectors ($\textbf{w}_k$) are chosen as the normalized versions of the combining vectors ($\textbf{v}_k$), as described in equation \eqref{w_k}. The computation of $||\textbf{v}_k||$ requires $2N$ real multiplications\footnote{{Consider the complex vector $\textbf{x} = [x_1,\dots,x_N]$. The computation of $||\textbf{x}|| = \sqrt{\sum\limits_{\substack{n=1}}^N |x_n|^2 }$ depends on previously obtaining $|x_m|^2$. Being $x_n = a_n + jb_n$ a complex scalar, $|x_n|^2 = a_n^2 + b_n^2$ requires 2 real multiplications and 1 real sum. Therefore, the computation of $||\textbf{x}_k||$ requires $2N$ real multiplications.}} and the division of $\textbf{v}_k$ by $||\textbf{v}_k||$ also requires $2N$ real divisions\footnote{{The division of a complex scalar $x = a + jb$ by a real scalar $c$ requires 2 real divisions. Therefore, the division of $\textbf{x}$ by $||\textbf{x}||$ requires $2N$ real divisions.}}, resulting in a total of $4N$ flops. Thus, the computation of the precoding matrix has a complexity of:}
\begin{equation}
	  {C_\text{SP-C}^\text{DL} = 4KN} \qquad\qquad {\text{[fpcb]}}
    \label{C-SP-DL}
	\end{equation}
	
{Notice that the rows of the combining and the precoding vectors of the user $k$ that correspond to the antennas whose indices are not in the set $\mathcal{D}_k$ are set equal zero. It reduces the complexity to obtain $y_k$, as in \eqref{receiveSP}, because the BS will use $N$ elements of the vectors $\textbf{v}_k$ and $\textbf{r}$, instead of $M$ elements, resulting in $3N$ flops. This procedure is repeated $\tau_\text{u}$ times per coherence block. Similarly, the complexity to precode the information during DL, $\textbf{w}_k x_k$, following \eqref{transmitSP}, is reduced since the BS only uses $N$ elements of the vector $\textbf{w}_k$, also resulting in $3N$ flops. This task is performed $\tau_\text{d}$ times per coherence block. Finally, the computational complexity associated to the reception and transmission of the information, in number of flops per coherence block, is:}
\begin{equation}
{C_\text{SP-R/T} = 3(\tau_\text{u}+\tau_\text{d})KN} \qquad\qquad {\text{[fpcb]}}
\label{C-SP-R/T}
\end{equation}

{To obtain $\theta_{mk}$ in \eqref{theta}, the BS computes $2K$ real multiplications and 1 real division. As there are $M$ antennas and $K$ users, the associated complexity is:}
\begin{align}
{C_{\text{SP},\theta} = (2K+1)MK} \qquad\qquad {\text{[fpcb]}}
\label{C-SP-theta}
\end{align}

{Finally, the {\it total signal processing computational complexity}, in flops per coherence block, when employing MR and ZF processing in the context of XL-MIMO antenna selection is given respectively by:}
\begin{equation}
{C_\text{TSP} = C_\text{SP-C}^\text{UL-MR} + C_\text{SP-C}^\text{DL} + C_\text{SP-R/T} + C_{\text{SP},\theta}} \,\,\, {\text{[fpcb]}}
\label{C-SP-MR}  \end{equation}
{and}
\begin{equation}
{C_\text{TSP} = C_\text{SP-C}^\text{UL-ZF} + C_\text{SP-C}^\text{DL} + C_\text{SP-R/T} + C_{\text{SP},\theta}} \quad {\text{[fpcb]}}
\label{C-SP-ZF}
	\end{equation}

{Hence, if antenna selection (AS) procedure is not applied, the complexities given in \eqref{C-SP-UL-ZF}, \eqref{C-SP-DL} and \eqref{C-SP-R/T} will be higher, since all the $M$ antennas always will be active for all the $K$ users, \textit{i.e.}, $N = M$.}

\section{SE and EE in XL-MIMO Systems}\label{sec:SE&EE}
	
The SE is defined as the sum-rate {in bits per channel use [bpcu]} achieved in the UL +  DL, expressed as:
\begin{equation}
	\text{SE} = \text{SE}_\text{UL} + \text{SE}_\text{DL} \qquad \text{{[bpcu]}}
	\label{SE}
	\end{equation}
	
{The overall network EE can be defined as the number of bits that can be reliably transmitted per unit of energy, which is the same as the throughput per unit of power $\left[\frac{\text{bit/s}}{\text{W}}\right]$, given by:}
\begin{equation}
\text{EE} = \frac{B\cdot\text{SE}}{P_\text{TX}^\text{UL} + P_\text{TX}^\text{DL} + P_\text{TX}^\text{tr} + P_\text{CP}}\qquad \left[\frac{\text{bit}}{\text{J}}\right] 
	\label{EE}
	\end{equation}
	\noindent where the denominator includes all power consumption terms required to make the wireless communication system operational. Hence, the term
	\begin{equation}
	P_\text{TX}^\text{tr} = \frac{\tau_\text{p}}{\tau_\text{c}}\frac{1}{\eta^\text{UL}}K p_\text{p}
	\label{P_TX_tr}
	\end{equation}
accounts for the total power consumed by the power amplifiers during the UL pilot transmission, while
	\begin{equation}
	P_\text{TX}^\text{UL} = \frac{\tau_\text{u}}{\tau_\text{c}}\frac{1}{\eta^\text{UL}} \sum_{k=1}^K p_k^\text{UL}
	\label{P_TX_UL}
	\end{equation}
and
	\begin{equation}
	P_\text{TX}^\text{DL} = \frac{\tau_\text{d}}{\tau_\text{c}}\frac{1}{\eta^\text{DL}} \sum_{k=1}^K p_k^\text{DL}
	\label{P_TX_DL}
	\end{equation}
	
\noindent refers to the UL and DL power consumed for data transmission, respectively, being $\eta^\text{UL}$ and $\eta^\text{DL}$ the power amplifier efficiency at the BS and at the users, respectively. $P_\text{CP}$ represents the circuit power consumption. A detailed model for $P_\text{CP}$ is discussed in the sequel.

\subsection{Circuit Power Model}\label{subsec:SE&EE_CPmodel}
The following circuit power consumption model based on \cite{massivemimobook} is adopted in this work:
	\begin{equation}
	P_\text{CP} = P_\text{FIX} + P_\text{TC} + P_\text{CE} + P_\text{C/D} + P_\text{BH} + P_\text{SP}
	\label{P_CP}
	\end{equation}
\noindent where $P_\text{FIX}$ is a constant quantity. It accounts for the power consumption required for site-cooling, control signaling and load-independent power of backhaul infrastructure and baseband processors. The power consumed by the backhaul is commonly modeled as the sum of two parts: load-independent and load-dependent. The last one will be included in $P_\text{BH}$, and is typically the least significant part (around 20\%) \cite{massivemimobook}.
	
The other terms of the model represented in \eqref{P_CP} account for the power consumption of the transceiver chains ($P_\text{TC}$), the channel estimation process ($P_\text{CE}$), the channel coding and decoding units ($P_\text{C/D}$), the load-dependent backhaul ($P_\text{BH}$) and the linear processing at the BS ($P_\text{LP}$). Each of these terms depends on at least one of the main system parameters: $M$, $K$ and the {ergodic spectral efficiency (SE)}.
	
The power consumption of the transceiver chains ($P_\text{TC}$) involves the power consumed by the BS local oscillator ($P_\text{LO}$), the power required by the circuit components (converters, mixers and filters) of each BS antenna ($P_\text{BS}$) and the power necessary to run the circuit components (mixers, filters, amplifiers and oscillator) of each single-antenna user ($P_\text{UE}$), as described by the equation
\begin{equation}
P_\text{TC} = P_\text{LO} + N_\text{act} P_\text{BS} + K P_\text{UE}
\label{P_TC}
\end{equation}

{The computational complexity associated to the channel estimation process is given by the equation \eqref{C_CE}. Hence, the resulting power consumption is given by\footnote{{As each coherence block contains $\tau_\text{c}$ symbols per second, $B/\tau_\text{c}$ is the number of coherence blocks per second. If a given signal processing has a computational complexity denoted by $C$, representing the number of flops per coherence block, and $L$ is the computational efficiency, representing the number of flops per Joule of energy, then $\frac{C}{L}$ represents the energy consumption per coherence block. Therefore, the associated power consumption is $\frac{BC}{\tau_\text{c}L}$.}}:}
\begin{equation}
{P_\text{CE} = \frac{B C_\text{CE}}{\tau_\text{c}L_\text{BS}}}
\label{P_CE}
\end{equation}
{where $L_\text{BS}$ is the computational efficiency at the BS, in [flop/s~W]. Similarly, the total signal power consumption is given by:}
\begin{equation}
{P_\text{SP} = \frac{B C_\text{TSP}}{\tau_\text{c}L_\text{BS}}}
	\label{P_SP}
\end{equation}

    The power consumed by the channel coding and decoding units is defined as
	\begin{equation}
	P_\text{C/D} = B~\text{SE}~( \cal{P}_\text{COD} + \cal{P}_\text{DEC} )
	\label{P_C/D}
	\end{equation}
    which increases linearly with the actual rates. $\mathcal{P}_\text{COD}$ and $\mathcal{P}_\text{DEC}$ are the coding and decoding power densities, respectively, in $\left[ \frac{\text{watt}}{\text{bit/s}} \right]$. For simplicity, $\mathcal{P}_\text{COD}$ and $\mathcal{P}_\text{DEC}$ are assumed to be the same in both UL and DL.
	
    The load-dependent backhaul power consumption, necessary for the UL and DL data transmission between the BS and the core network, is modeled as
	\begin{equation}
	P_\text{BH} = B~\text{SE}~\cal{P}_\text{BT}
	\label{P_BH}
	\end{equation}
    where $\cal{P}_\text{BT}$ is the backhaul traffic power density, in $\left[ \frac{\text{watt}}{\text{bit/s}} \right]$. There is also a load-independent backhaul power consumption, which can be included in $P_\text{FIX}$.


\section{Numerical Results}
	\label{sec:results}
{In the sequel we present numerical results based on Monte-Carlo simulations with the objective to demonstrate that the proposed algorithm provides an EE increase while reducing considerably the computational complexity and the power consumption in the context of XL-MIMO systems. Table \ref{tab:parameters} contains a list of the main deployed parameter values,} similar to those adopted in \cite{massivemimobook}, \cite{BJORNSON2015}, \cite{marinelloEE}.
	
	\subsection{{Simulation Setup and System Configuration}}
	\label{sec:results_part1}
	
	{In our simulations, the antenna array contains $M$ = 100 antenna elements and each user sees the array through 3 different VRs, \textit{i.e.}, $N_\text{c}=3$. Furthermore, $N_{ik}$ is taken as a uniform random variable distribution over the interval $[0.1M,0.3M]$, while the index of the first antenna that is inside this VR, denoted by $c_{ik}$, follows a uniform distribution in the interval $[1,M-N_{ik}]$.}
	
{The set $\mathcal{C}_k = \mathcal{C}_{1,k}\ \cup\ \dots\ \cup\ \mathcal{C}_{N_\text{C},k}$ contains the indices of the antennas that sees the $k$-th user. Thus, the number of elements in $\mathcal{C}_k$ corresponds to the total number of antennas seen by user $k$. We obtained this value numerically from our simulations. As well as the other numerical results presented in the paper, this value was averaged out of 1,000 random realizations, obtaining the average number of active antennas for the $k$-user as $|\mathcal{C}| = 55.8$. It means that approximately 56 antennas are seen by each user, on average. Also from our simulations, we found that, when $K$ = 4, each antenna sees an average of 2.23 users, while for $K$ = 40 this number goes to 22.3, resulting in higher interference levels.}
	
{Moreover, we assume equal power allocation (EPA), what means the transmit power is the same for all users. Hence, during the UL data transmission, $p_k^\text{UL} = p_\text{UL}$, while during the DL data transmission, the total available transmit power at the BS, $P_\text{DL}$, is equally divided among the $K$ users, so that the power allocated to the $k$-th user is $p_k^\text{DL} = P_\text{DL}/K$. We have adopted $p_\text{P}=p_\text{UL}$ = 0.1 W and $P_\text{DL}$ = 1 W. Although the position of the users (and consequently the pathloss), the short-scale fading and the VRs are random variables, the numerical results are statically relevant because they represent an average over 1,000 realizations.}
	
\begin{table}[!htbp]
\caption{{Summary of system and channel adopted parameter values, similar to those adopted in \cite{massivemimobook}, \cite{BJORNSON2015}, \cite{marinelloEE}}}
\begin{center}
\resizebox{.58\textwidth}{!}{
\begin{tabular}{|l|c|} \hline
	\textbf{Parameter} & \textbf{Value} \\ \hline \hline
	Pathloss attenuation exponent: $\gamma$ & 2.5\\ \hline
	Median channel gain at a distance of 1 m: $b_0$ & $2.95\cdot10^{-4}$\\ \hline
	Number of NLoS VRs for each user: $N_\text{C}$ & 3\\ \hline
	Antenna array length: $L$ & 60 m \\ \hline
	Number of BS antennas (ULA), $M$ & $100$\\ \hline
	Number of mobile users, $K$ & $\{4; \,\, 40\}$\\ \hline
	Minimum distance ($d_\text{min}$) & 5 m \\ \hline
	Maximum distance ($d_\text{max}$) & 30 m \\ \hline
	Number of Monte Carlo realizations: $\mathcal{T}$ & 1,000 \\ \hline
	Transmission bandwidth: $B$ & 20 MHz\\ \hline
	Channel coherence bandwidth: $B_\text{C}$ & 100 kHz\\ \hline
	Channel coherence time: $T_\text{C}$ & 2 ms\\ \hline
	Total UL noise power: $\sigma_\text{UL}^2$ & $-$ 100 dBm\\ \hline
	Total DL noise power: $\sigma_\text{DL}^2$ & $-$ 80 dBm\\ \hline
    UL pilot transmit power per user: $p_\text{P}$ & 0.1 W\\ \hline
    UL data transmit power per user: $p_\text{UL}$ & 0.1 W\\ \hline
    Total DL data transmit power: $P_\text{DL}$ & 1.0 W\\ \hline
    Fraction of UL transmission: $\epsilon_\text{u}$ & 0.4\\ \hline
    Fraction of DL transmission: $\epsilon_\text{d}$ & 0.6\\ \hline
    Power amplifier efficiency at the users: $\eta^\text{UL}$ & 0.5\\ \hline
    Power amplifier efficiency at the BSs: $\eta^\text{DL}$ & 0.4\\ \hline
    Computational efficiency at the BS: $L_\text{BS}$ & 75 $\left[ \frac{\text{Gflop/s}}{\text{W}} \right]$\\ \hline
    Fixed power consumption: $P_\text{FIX}$ & 10 W\\ \hline
    Power consumed by local oscillators at BS: $P_\text{SYN}$ & 0.2 W\\ \hline
    Power consumed by circuit components at BS: $P_\text{BS}$ & 0.2 W\\ \hline
	Power consumed by circuit components  at UE: $P_\text{UE}$ & 0.2 W\\ \hline
	Power density for coding of data signals: $\cal{P}_\text{COD}$ & 0.1 $\left[ \frac{\text{W}}{\text{Gbit/s}} \right]$\\ \hline
	Power density for decoding of data signals: $\cal{P}_\text{DEC}$ & 0.8 $\left[ \frac{\text{W}}{\text{Gbit/s}} \right]$\\ \hline
	Power density for backhaul traffic: $\cal{P}_\text{BT}$ & 0.25 $\left[ \frac{\text{W}}{\text{Gbit/s}} \right]$\\ \hline
\end{tabular}
}
\end{center}
\label{tab:parameters}
\end{table}

\begin{figure}[!htbp]
\centering
\includegraphics[width=.65\textwidth]{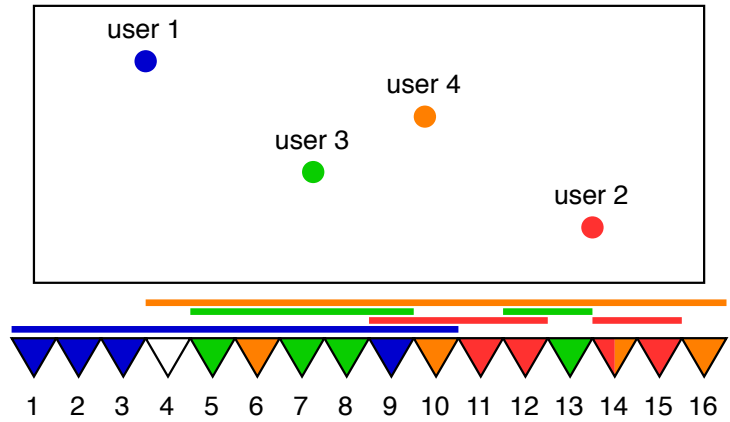}
\caption{{Example of a XL-MIMO system, with 4 users being served by a BS equipped with a 16-antenna linear antenna array. The figure illustrates each user's VR and the antennas that the proposed algorithm designates to communicate with each user.}}
\label{fig:example}
\end{figure}

{Figure \ref{fig:example} shows an example of the XL-MIMO system user spatial distribution, where $K=4$ and $M=16$, for simplicity. Each triangle represents one of the 16 BS antennas, while each colored circle represents one of the 4 mobile users, which are randomly distributed over a rectangular area. Taking a random channel realization, the portion of the array that each user sees is indicated by the horizontal line with its correspondent color. User 1, for example, sees the antennas 1 to 10, while user 2 sees the antennas 9 to 15, excepting the antenna 13. This fragmentation of the VR into two parts may occur if any object is blocking the signal in that region.}
    
{In this example, the algorithm was set up to define $N=4$ antennas to communicate with each user. In the figure, the triangle that represents a given antenna $m$ is painted with the user $k$ color if user $k$ sees antenna $m$. For example, antennas 1, 2, 3 and 9, which are in blue, were designated to communicate with user 1. Notice that all these 4 antennas are part of the VR of user 1. Although the signal from user 1 probably achieves the antennas 4 to 8 with higher intensity than antenna 9, choosing one of these antennas would increase the received interference power, mainly due to user 3. Observe that antenna 4 is not active, while antenna 14 serves users 2 and 4, simultaneously. Therefore, considering that only one antenna was designated to communicate with 2 users at the same time, we can say that the algorithm succeeded in avoiding the interference from other users to affect the SINR while reducing the computational complexity.}
    
\subsection{{Dependency of the System Performance on $N$}}
\label{sec:results:part2}
\begin{figure*}[htbp!]
		\centering
		\begin{subfigure}{0.5\textwidth}
			\centering
		\includegraphics[width=\textwidth]{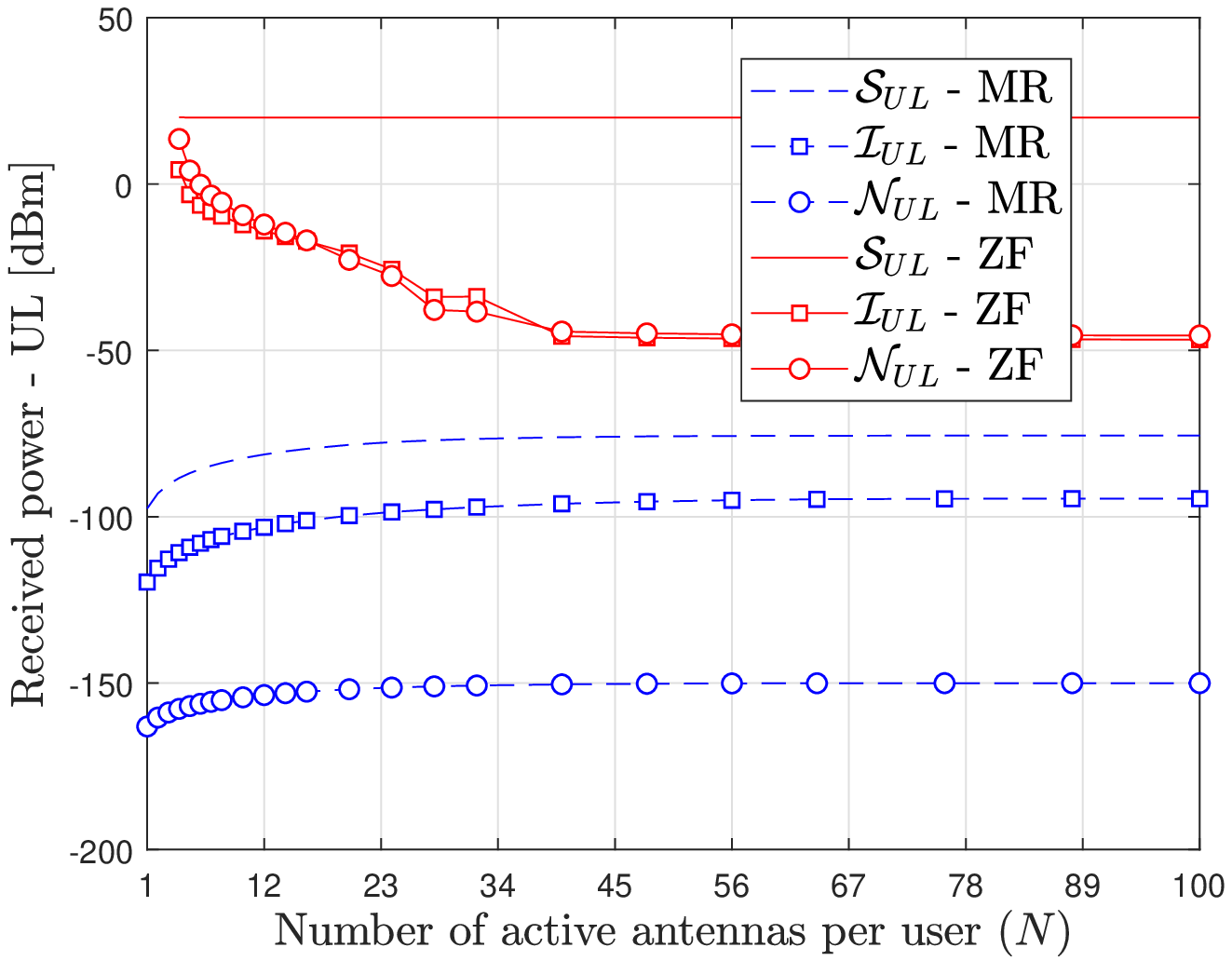}
			\caption{UL received power, $K$ = 4}
			\label{K04_RecPowerUL}
		\end{subfigure}%
		\begin{subfigure}{0.5\textwidth}
			\centering
			\includegraphics[width=\textwidth]{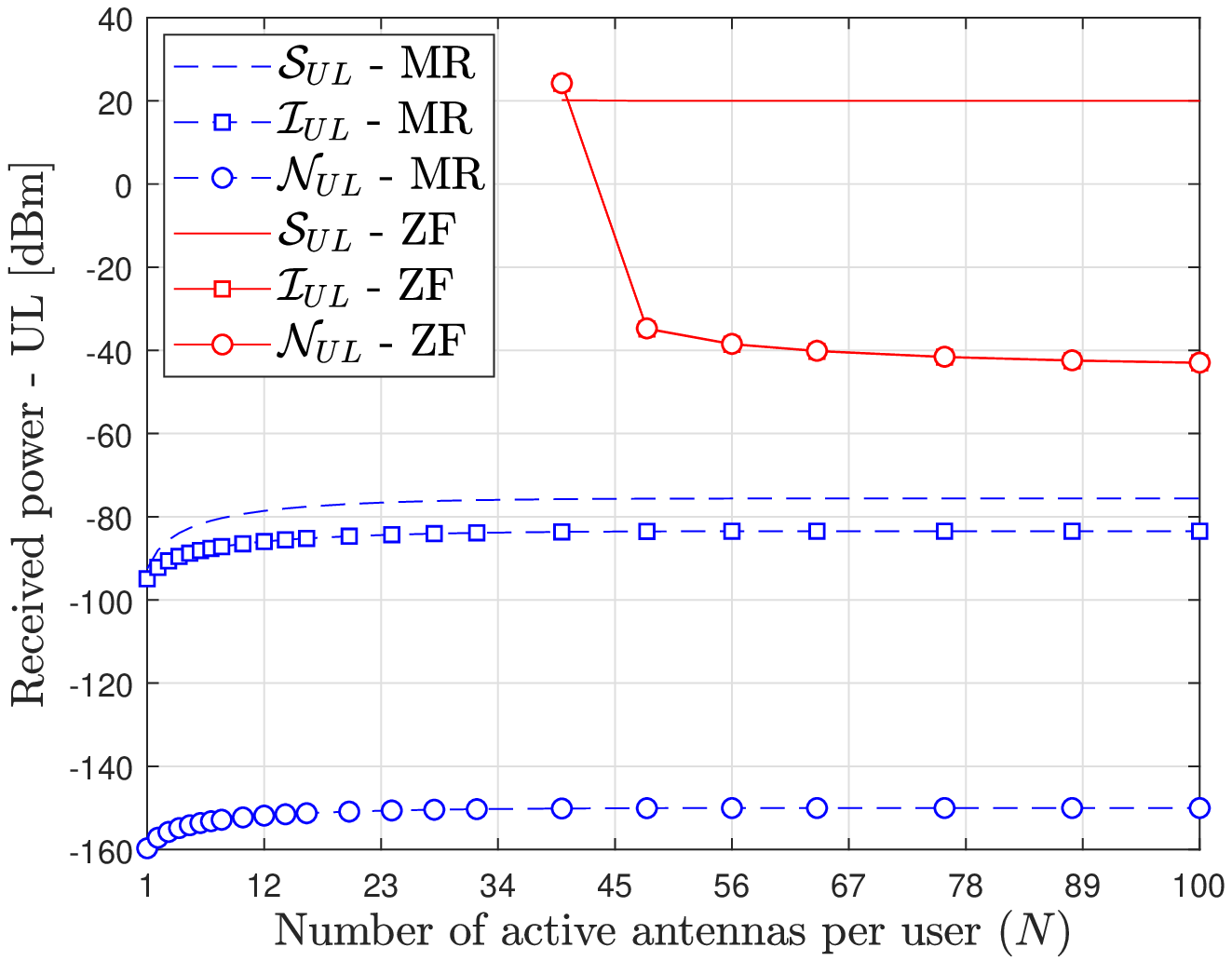}
			\caption{UL received power, $K$ = 40}
			\label{K40_RecPowerUL}
		\end{subfigure}
		\newline
		\begin{subfigure}{0.5\textwidth}
			\centering
			\includegraphics[width=\textwidth]{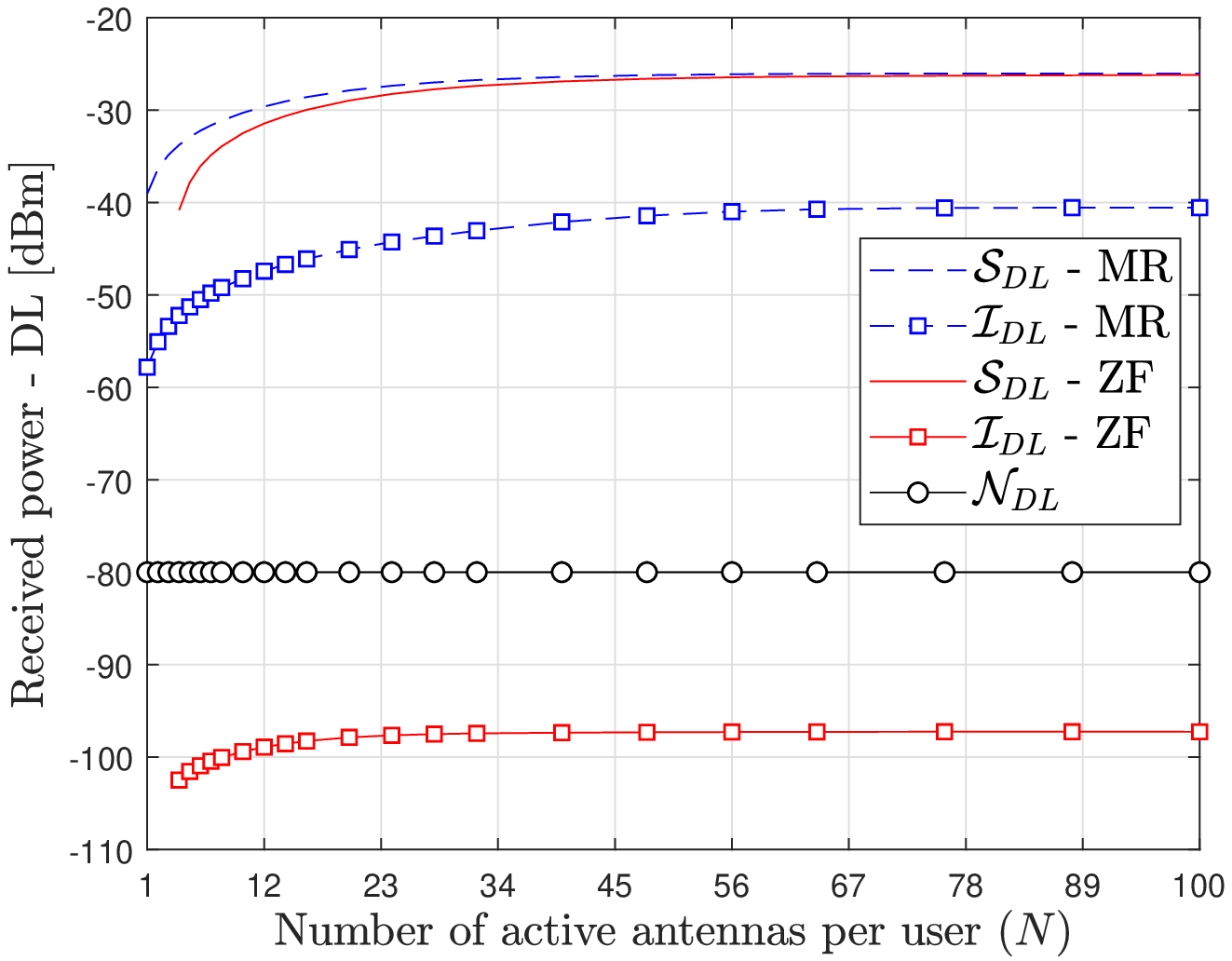}
			\caption{DL received power, $K$ = 4}
			\label{K04_RecPowerDL}
		\end{subfigure}%
		\begin{subfigure}{0.5\textwidth}
			\centering
			\includegraphics[width=\textwidth]{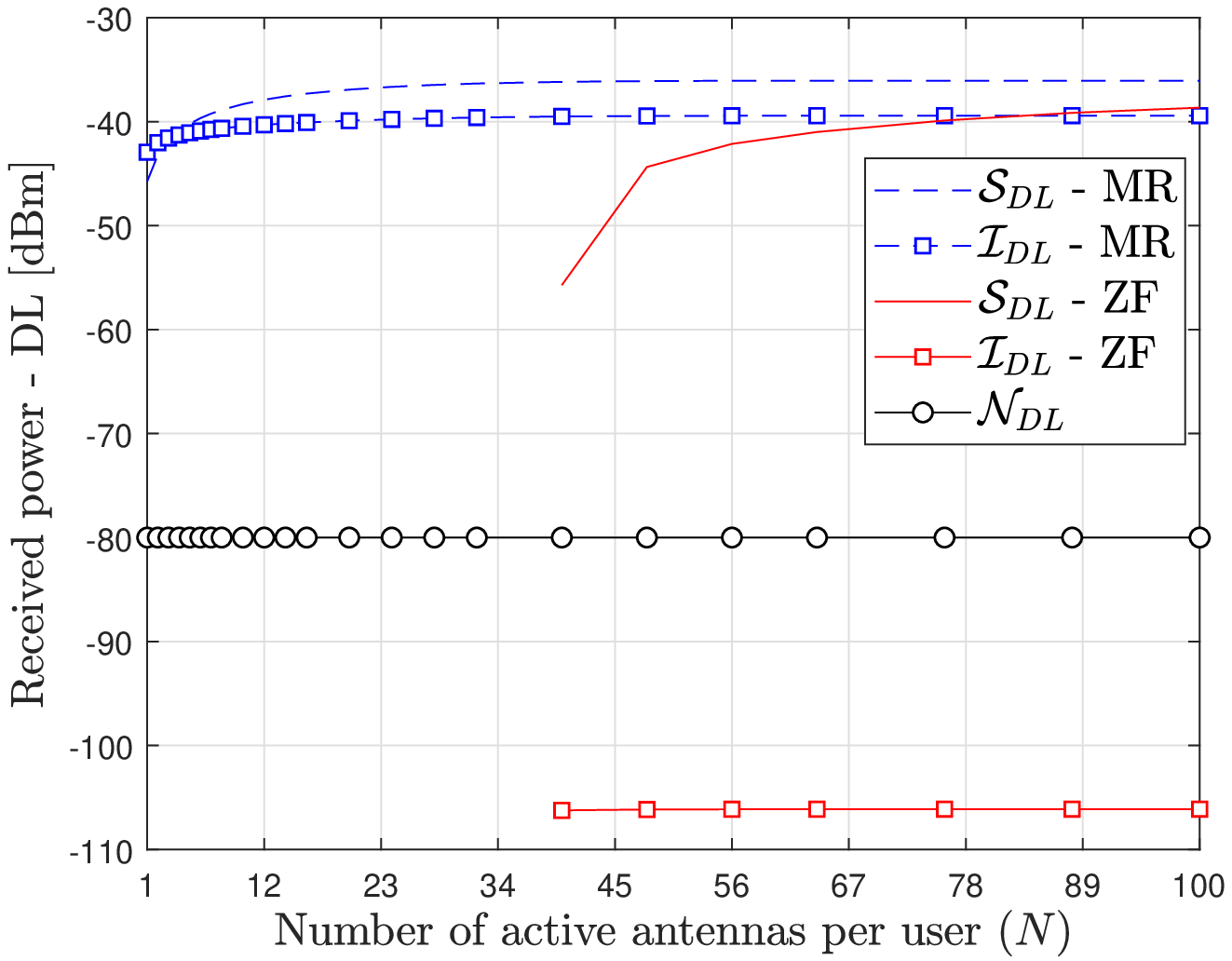}
			\caption{DL received power, $K$ = 40}
			\label{K40_RecPowerDL}
		\end{subfigure}
		\newline
        \begin{subfigure}{0.5\textwidth}
			\centering
		\includegraphics[width=\textwidth]{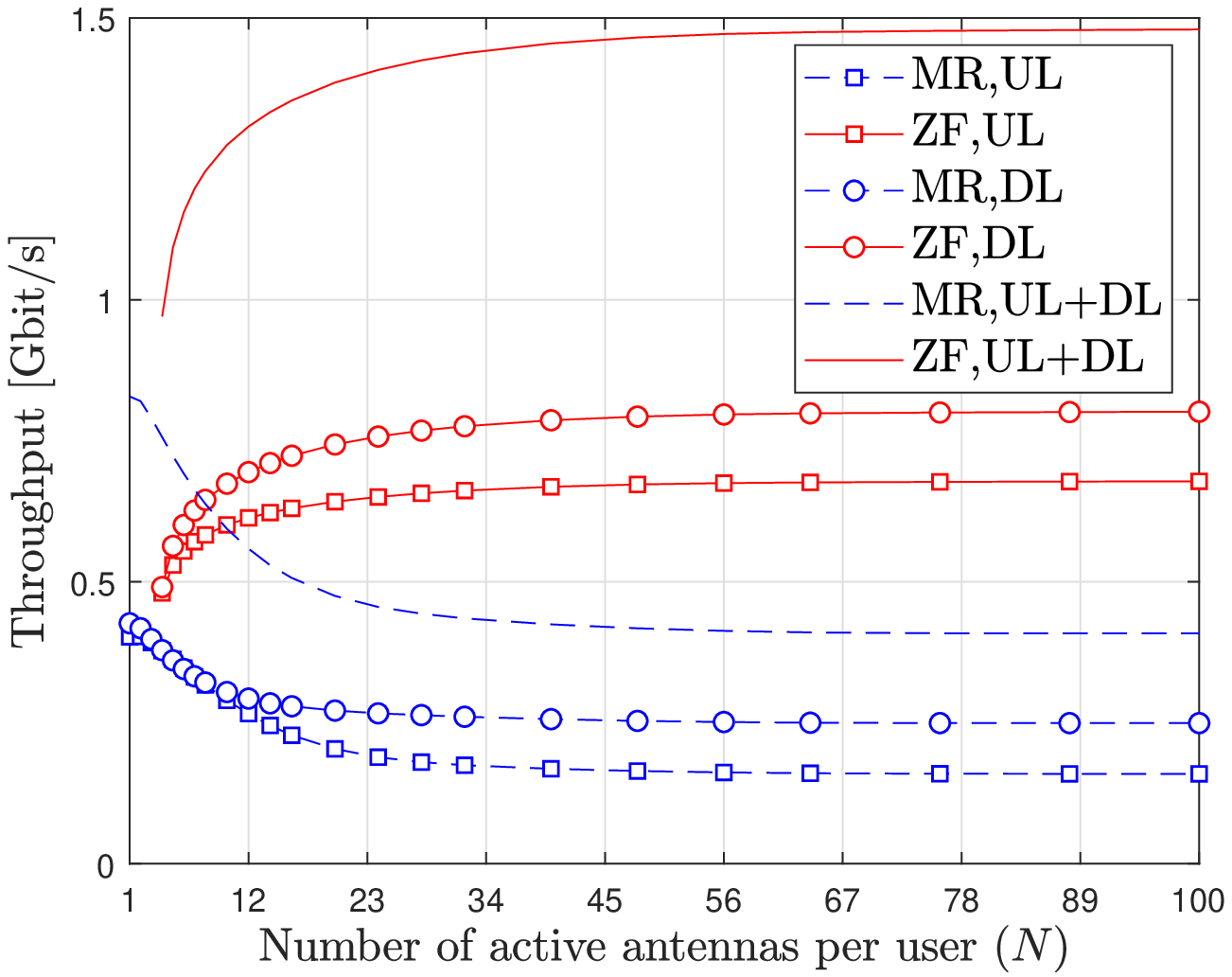}
			\caption{Throughput, $K$ = 4}
			\label{K04_SE}
		\end{subfigure}%
		\begin{subfigure}{0.5\textwidth}
			\centering
		\includegraphics[width=\textwidth]{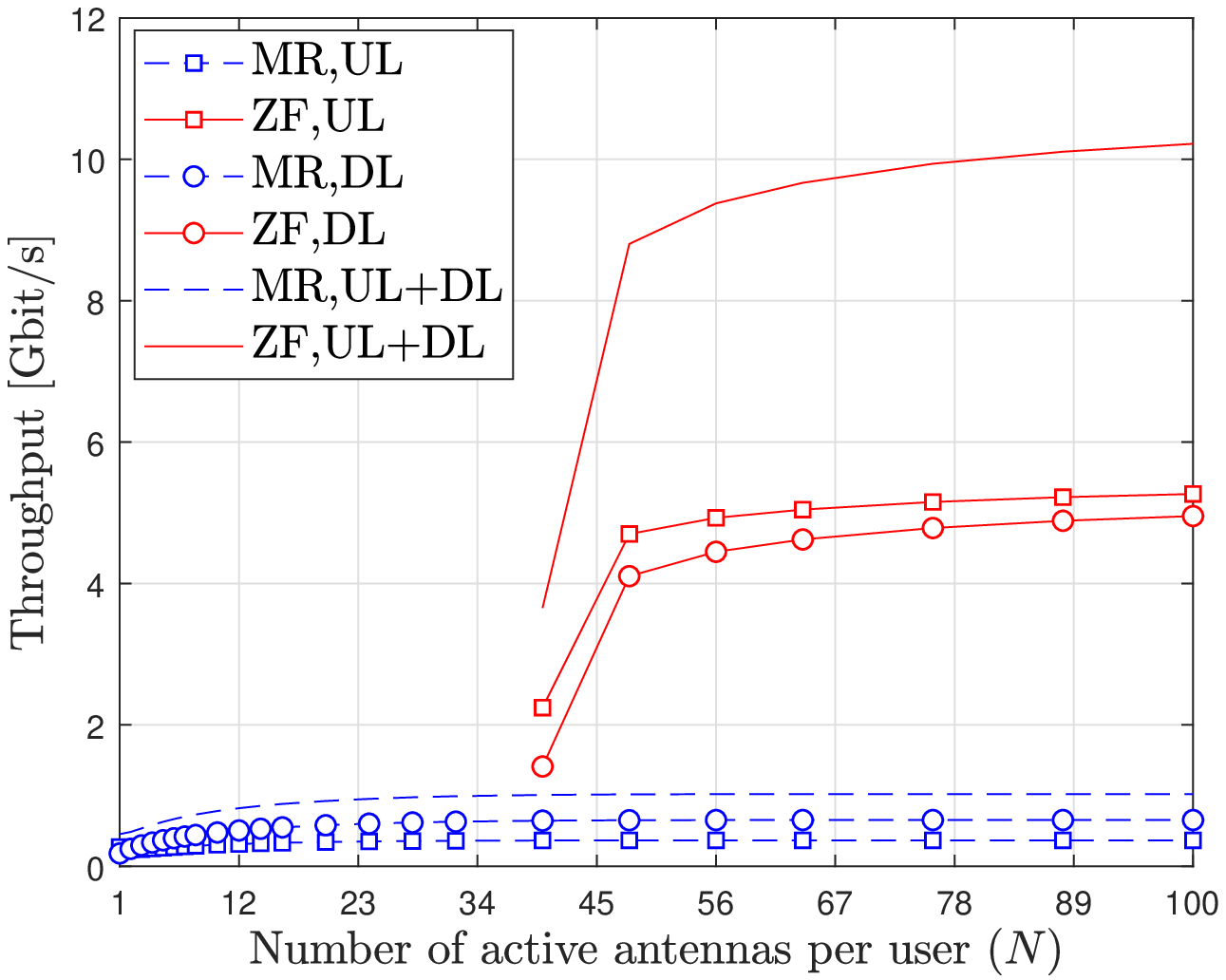}
			\caption{Throughput, $K$ = 40}
			\label{K40_SE}
		\end{subfigure}
	\caption{{Average received power of the desired signal, undesired signal (interference) and noise as given by Def. 1 to 6), as well as the system throughput, during the UL and the DL data transmission. Two scenarios are considered: $K$ = 4 and $K$ = 40 users.}}
		\label{fig:S_I_N_throughput}
	\end{figure*}

{In the following, we present numerical results demonstrating how the number of active antennas per user ($N$) influences the system performance, in terms of throughput and EE, and the resource consumption, in terms of computational complexity and power consumption. From these results, we can see the advantages of appropriately selecting the antennas subset (AS strategy) against utilizing the whole antenna array strategy to serve all users at the same time (no-AS strategy). To refine the comprehension upon these results, which depend on the processing scheme (MR or ZF), Fig. \ref{fig:S_I_N_throughput} provides, for two different scenarios ($K$ = 4 and $K$ = 40), valuable insights on how $N$ influences the received signal, interference and noise power, during the UL and during the DL, which are given by the definitions in Table \ref{tab:defs}.} 
    
\begin{table}[!htbp]
\caption{{Definitions considered in Fig. \ref{fig:S_I_N_throughput}, in [dBm].}}
\begin{center}
\def\arraystretch{1.5} 
\begin{tabular}{l} \hline
    {\textbf{Definition 1:} UL average received signal power}: \\ ${\mathcal{S}_\text{UL} = 10\cdot \log_{10} \,\left(\frac{1}{K}\sum\limits_{\substack{k=1}}^K \mathbb{E} \{p_k^\text{UL}|\textbf{v}_k^\text{H}\textbf{h}_k|^2\}\right) + 30}$ \\ \hline
    {\textbf{Definition 2:} UL average received interference power}\\ ${\mathcal{I}_\text{UL} = 10\cdot \log_{10} \,\left(\frac{1}{K}\sum\limits_{\substack{k=1}}^K \sum\limits_{\substack{i=1\\i\neq k}}^K \mathbb{E}\{p_i^\text{UL}|\textbf{v}_k^\text{H}\textbf{h}_i|^2\}\right) + 30}$ \\ \hline
    {\textbf{Definition 3:} UL average received noise power} \\ ${\mathcal{N}_\text{UL} = 10\cdot \log_{10} \,\left(\frac{\sigma_\text{UL}^2}{K}\sum\limits_{\substack{k=1}}^K \mathbb{E}\{||\textbf{v}_k||^2\}\right) + 30}$ \\ \hline\hline
    {\textbf{Definition 4:} DL average received signal power} \\ ${\mathcal{S}_\text{DL} = 10\cdot \log_{10} \,\left(\frac{1}{K}\sum\limits_{\substack{k=1}}^K \mathbb{E} \{p_k^\text{DL}|\textbf{h}_k^\text{H}\textbf{w}_k|^2\}\right) + 30}$ \\ \hline
    {\textbf{Definition 5:} DL average received interference power} \\ ${\mathcal{I}_\text{DL} = 10\cdot \log_{10} \,\left(\frac{1}{K}\sum\limits_{\substack{k=1}}^K \sum\limits_{\substack{i=1\\i\neq k}}^K \mathbb{E}\{p_i^\text{DL}|\textbf{h}_k^\text{H}\textbf{w}_i|^2\}\right) + 30}$ \\ \hline
    {\textbf{Definition 6:} DL average received noise power} \\ ${\mathcal{N}_\text{DL} = 10\cdot \log_{10} \,\left(\sigma_\text{DL}^2\right) + 30}$ \\ \hline
    \end{tabular}
    \end{center}
    \label{tab:defs}
    \end{table}

{Figures \ref{K04_RecPowerUL} and \ref{K40_RecPowerUL} compare the UL signal, interference and noise power, as given by the definitions Def. 1 to 3, when the number of users is 4 and 40, respectively. When $N < K$, it is not possible to execute ZF, due to singularity problems with the matrix $\hat{\textbf{H}}^\text{H} \hat{\textbf{H}}$. First, one can see that the received power levels of ZF are higher than of MR. The reason is that ZF receive combining is the pseudoinverse of $\hat{\textbf{H}}$ (recall that the elements of the channel matrix include the pathloss effect), while MR combining is simply a copy of $\hat{\textbf{H}}$. Second, by comparing Figures \ref{K04_RecPowerUL} and \ref{K40_RecPowerUL}, one can observe that,as expected,  when $K$ = 40, the received interference power ($\mathcal{I}_\text{UL}$) is higher than when $K$ = 4, because the selectivity of the receive combining deteriorates. Third, looking at the MR curves, one can verify that, by selecting more antennas to be active for each user, the received signal power increases, but so does the interference and noise power. It does not occur when employing ZF combining. By increasing $N$, ZF performs better at reducing the interference and noise power, because the receive combining becomes more selective. Fourth, if the channel estimates are reliable, ZF combining will force the average received signal power ($\mathcal{S}_\text{UL}$) to $p_\text{UL}$, as $|\textbf{v}_k^\text{H}\textbf{h}_k|^2 = 1$, from the definition of ZF combining. It can be observed in the figure, what attests that the channel estimation predicted by eq. \eqref{h_est} provides good estimates.}
    
{Unlike the UL case, during the DL (see Fig. \ref{K04_RecPowerDL} and \ref{K40_RecPowerDL}), the average received noise, $\mathcal{N}_\text{DL}$, does not depend on the precoding scheme (compare Def. 3 and 6). Another important point is that the desired signal power ($\mathcal{S}_\text{DL}$) is the same magnitude order for both MR and ZF, because the precoding vectors are normalized, unlike the combining vectors. Finally, as the total available DL transmit power ($P_\text{DL}$) is distributed among the $K$ users, the individual DL transmit power ($p_k^\text{DL}$) is inversely proportional to $K$. That is why the average signal and interference power are smaller in Fig. \ref{K40_RecPowerDL} than in Fig. \ref{K04_RecPowerDL}. The exception is $\mathcal{I}_\text{DL}$ when employing MR precoding, which is less efficient than ZF at eliminating the interference.}

{Fig. \ref{K04_SE} and \ref{K40_SE} depict the UL, DL and UL+DL throughput, given by $B\cdot\text{SE}_\text{UL}$, $B\cdot\text{SE}_\text{DL}$ and $B\cdot\text{SE}$, respectively. The throughput resulting from the no-AS strategy can be obtained in the point $N$ = 100. Thus, we can see that in a scenario with few active users, the AS strategy improves considerably the system throughput when using MR. On the other side, considering the $K$ = 40 scenario, the high interference levels deteriorate the selectivity of MR combining and precoding, and the AS strategy cannot improve the throughput. The same behavior is observed with ZF processing, independently of the number of users. However, as the array is physically large and the users have a VR corresponding to around 50\% of the array, on average (as discussed in the beginning of Section \ref{sec:results_part1}), increasing $N$ beyond $M/2$ can barely improves the throughput. Thus, by taking $N$ close to $M/2$, we can obtain almost the same throughput achieved with the no-AS strategy, as shown in Figures \ref{K04_SE} and \ref{K40_SE}, while benefiting from lower computational complexity and consequently lower power consumption.} 

\subsection{Computational Complexity}

\begin{figure}[!htbp]
\centering
\includegraphics[trim={5mm 1mm 7mm 7mm},clip,width=.67\textwidth]{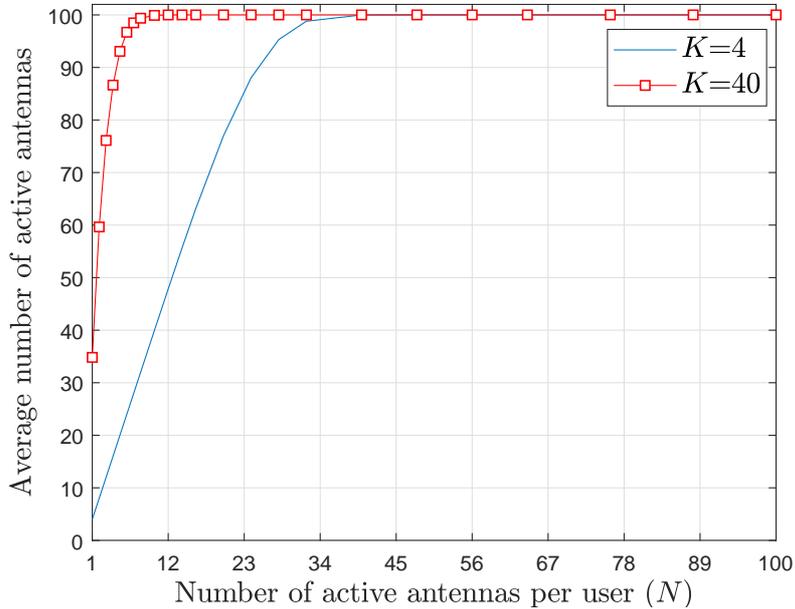}
	\caption{{Average number of active antennas per user ($N$) \textit{versus} the total number of active BS antennas ($N_\text{act}$).}}
	\label{active_antennas}
\end{figure}

\begin{figure*}[!htbp]
	\centering
	\begin{subfigure}{0.5\textwidth}
		\centering
		\includegraphics[width=\textwidth]{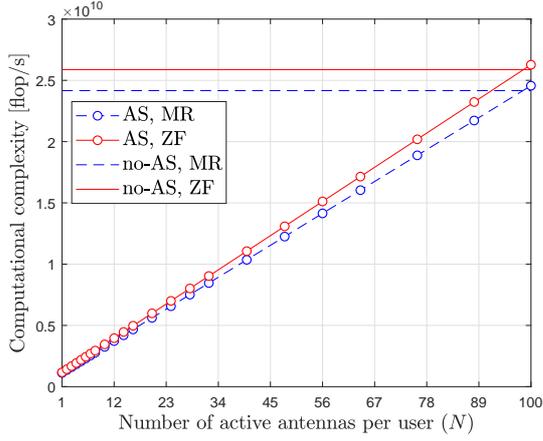}
		\caption{Computational complexity, $K$ = 4}
		\label{K04_C}
	\end{subfigure}%
	\begin{subfigure}{0.5\textwidth}
		\centering
		\includegraphics[width=\textwidth]{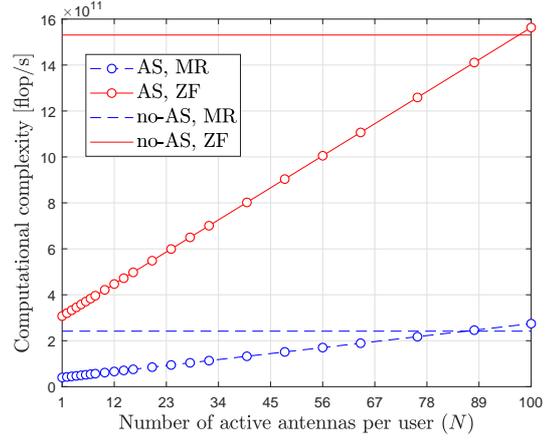}
		\caption{Computational complexity, $K$ = 40}
		\label{K40_C}
	\end{subfigure}%
	\newline
	\begin{subfigure}{0.5\textwidth}
		\centering
		\includegraphics[width=\textwidth]{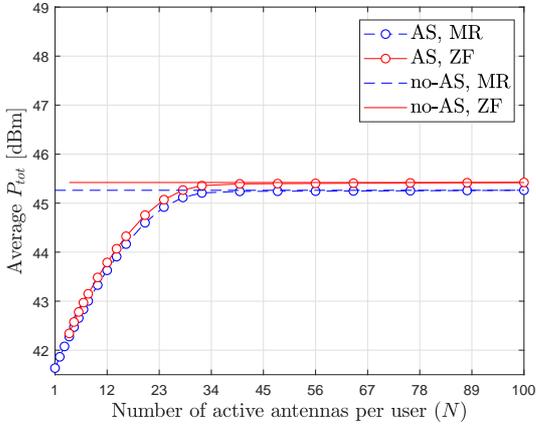}
    	\caption{Power consumption, $K$ = 4}
		\label{K04_P}
	\end{subfigure}%
	\begin{subfigure}{0.5\textwidth}
		\centering
		\includegraphics[width=\textwidth]{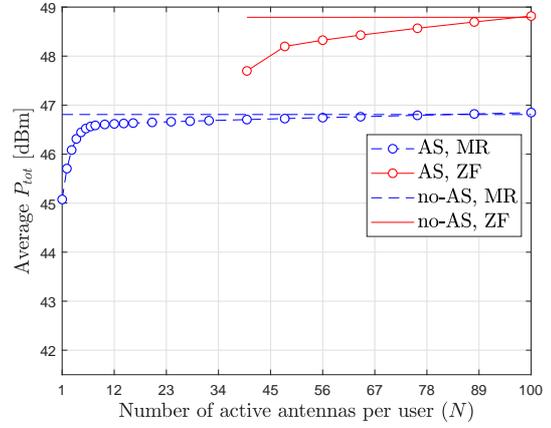}
		\caption{Power consumption, $K$ = 40}
		\label{K40_P}
	\end{subfigure}%
	\caption{{The average consumption of computational and power resources raises when increasing $N$, while depends on the adopted AS strategy.}}
	\label{fig:C&P}
\end{figure*}

{While $\mathcal{D}_k$ contains the indices of the $N$ antennas that are active for user $k$, the set $\mathcal{D}$ contains the indices of antennas $N_\text{act}$ that are active for any of the $K$ users. Fig. \ref{active_antennas} addresses the dependency of $N_\text{act}$ on $N$. In the scenario with 4 users, the whole antenna array is expected to be active when $N\geq34$, approximately. When $K$ = 40, $N_\text{act}$ scales faster than when $K$ = 4, and reaches the limit of 100 when $N\approx 8$. If a small $N$ is sufficient to substantially increase the EE or throughput, one will benefit from the reduced circuit power consumption, as $P_\text{TC}$ is proportional to $N_\text{act}$, according to \eqref{P_TC}. It can be evidenced by Fig. \ref{K04_P} and \ref{K40_P} that with small values of $N$, the AS strategy provides considerable reduction on the power consumption comparing to the entire antenna array activation (no-AS strategy).}
	
{Table \ref{tab:comp} provides a quantitative analysis of the impact of the proposed HRNP-based AS scheme on the overall XL-MIMO computational complexity, given by $C = C_\text{CE} + C_\text{TSP}$. The complexity associated to the no-AS strategy can be obtained from \eqref{C-SP-UL-ZF}, \eqref{C-SP-UL-MR}, \eqref{C-SP-DL} and \eqref{C-SP-R/T} by simply replacing $N$ with $M$ in these equations, and cancelling $C_{\text{SP},\theta}$ out in \eqref{C-SP-MR} and \eqref{C-SP-ZF}. The complexities are in the unit of [flop/s], which is given by $BC/\tau_\text{c}$, recalling that $C$ is measured in flop per coherence block [fpcb]. The first three columns of the table define four scenarios, with different values of $M$, $N$ and $K$. The other columns contain the complexity associated to each processing scheme (MR and ZF), when using or not the AS algorithm. Notice that the algorithm yields very lower complexities, which is particularly advantageous in high system dimensions (many BS antennas and many users). Also, although providing lower throughput than ZF, one can benefit from using MR due to its much smaller complexity. However, in the fourth scenario, the AS's complexity surpasses the no-AS's when employing MR, because $C_{\text{SP},\theta}$ is the only term of $C_\text{TSP}$ that is proportional to $K^2$. It does not occur with ZF, as the term $C_\text{SP-C}^\text{UL-ZF}$ is much more significative than $C_{\text{SP},\theta}$.}
	
\begin{table}[!htbp]
\caption{{AS computational complexity in [Gflop/s], discriminated by processing scheme.}}
\begin{center}
\resizebox{.63\textwidth}{!}{
\begin{tabular}{|r|r|r|r|r|r|r|} \hline
$M$ & $N$ & $K$ & no-AS MR & AS MR & no-AS ZF & AS ZF \\ \hline
  32 &   4 &   2 &      4 &      1 &       4 &       1 \\ \hline
 128 &  16 &   8 &     62 &     12 &      76 &      14 \\ \hline
 512 &  64 &  32 &    990 &    369 &   3,848 &     819 \\ \hline
2048 & 256 & 128 & 15,834 & 17,551 & 702,031 & 126,822 \\ \hline
\end{tabular}
}
\end{center}
\label{tab:comp}
\end{table}

{Fig. \ref{K04_C} shows the linear dependency of the computational complexity on $N$ and the remarkable computational complexity reduction provided by the adopted AS strategy. As a consequence, there is also a reduction in the total power consumption, defined as $P_\text{tot}=P_\text{TX}^\text{UL} + P_\text{TX}^\text{DL} + P_\text{TX}^\text{tr} + P_\text{CP}$, as depicted in Fig. \ref{K04_P}. Moreover, Fig. \ref{K40_C} reveals a significant complexity increase when the number of active users grows from 4 to 40. Furthermore, by avoiding all antennas to be simultaneously active, the adopted HRNP-AS strategy in eq. \eqref{theta} reduces the transceiver chains power consumption\footnote{Notice that from \eqref{P_TC}, power consumption is linearly dependent on $N_\text{act}$.}, and consequently the total power consumption, as indicated in Fig. \ref{K04_P} and \ref{K40_P}. As a final remark on the advantage in adopting the HRNP-AS strategy is the operation point where the AS power consumption curves meet the no-AS curves is very close to the point where the average number of active antennas meet $M$ in Fig. \ref{active_antennas}.}

\subsection{{Energy Efficiency}}
\begin{figure*}[!htbp]
		\centering
		\begin{subfigure}{0.5\textwidth}
\centering
		\includegraphics[width=\textwidth]{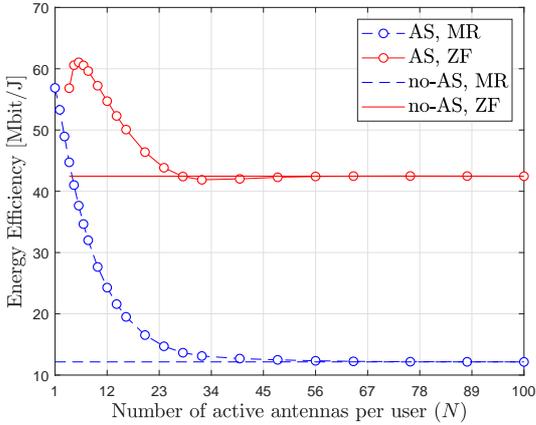}
			\caption{$K$ = 4}
			\label{K04_EE}
		\end{subfigure}%
		\begin{subfigure}{0.5\textwidth}
			\centering
\includegraphics[width=\textwidth]{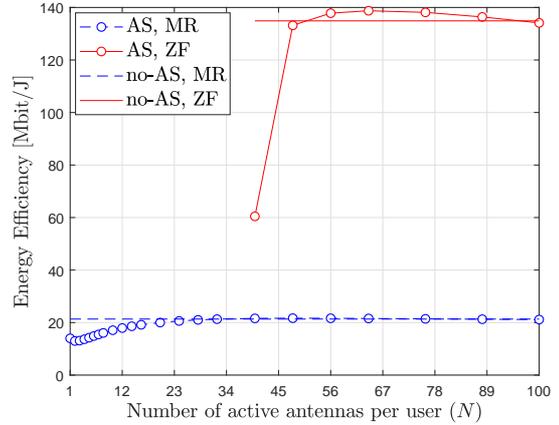}
\caption{$K$ = 40}
\label{K40_EE}
\end{subfigure}%
\caption{{Energy efficiency as a function of $N$, in both low and high loading scenarios: $K$ = 4 and $K$ = 40.}}
\label{fig:EE}
\end{figure*}

{Fig. \ref{fig:EE} confirms the EE improvement when AS strategy based on highest received power is adopted. Under reduced loading scenario ($K=4$), the EE is maximized when $N=6$ and $N=1$ for ZF and MR, respectively. It demonstrates that the HRNP-AS strategy can guarantee huge EE improvements in a scenario with few users, while reducing considerably the power consumption and the computational burden, as discussed above. On the other hand, results in Fig. \ref{K40_EE} demonstrate that, the AS strategy cannot improve the EE considerably when XL-MIMO operates under high loading scenarios ($K=40$). However, the HRNP-based AS strategy is still advantageous over the no-AS strategy, as it still provides a considerable complexity reduction. Thus, regardless of the number of users, it is not reasonable to use the whole antenna array to serve all users if about half of the antennas is sufficient to achieve a remarkable EE increasing with a lower computational complexity.}

\subsection{{Overall XL-MIMO Performance Comparison}}
\label{sec:results_part3}
{Table \ref{tab:N*xK} shows the optimal number of selected antennas ($N^*$) that maximizes the EE, when employing both linear MR or ZF filtering. According to the data, $N^*$ is strongly influenced by the number of active users $K$. As a rule of thumb, when the number of active users is up to a limit, namely $K_\text{max}$, the EE is maximized when the AS algorithm selects  less than $10\%$ of the antenna array to communicate with each user, \textit{i.e.}, $N^*\leq M/10$. From the data in Table \ref{tab:N*xK}, we see that $K_\text{max}=20$ for MR and $K_\text{max}=6$ for ZF.}
	
\begin{table}[!htbp]
\caption{{Optimal number of selected antennas for maximizing the EE ($N^*$) \textit{versus} the number of active users.}}
\begin{center}
\resizebox{.63\textwidth}{!}{
\begin{tabular}{|c|c||r|r|r|r|r|r|r|r|r|r|r|r|} \cline{2-14}
  \multicolumn{1}{c|}{} & $K$ & 2 & 4 & \bf 6 & 8 & 10 & 12 & 16 & \bf 20 & 24 & 32 & 40 & 50 \\ \hline\hline
  \multirow{2}{*}{$N^*$} & MR & 3 & 1 & 1 & 1 & 1 &  1 & 1 & \bf 1 & 48 & 48 & 48 & 47 \\ \cline{2-14}
  & ZF & 5 & 6 & \bf 8 &64 &62 &60 &58 &57 &57 &60 &65 &75 \\ \hline
\end{tabular}
}
\end{center}
\label{tab:N*xK}
\end{table}

\begin{figure*}[htbp!]
	\centering
	\begin{subfigure}{.5\textwidth}
		\centering
		\includegraphics[width=\textwidth]{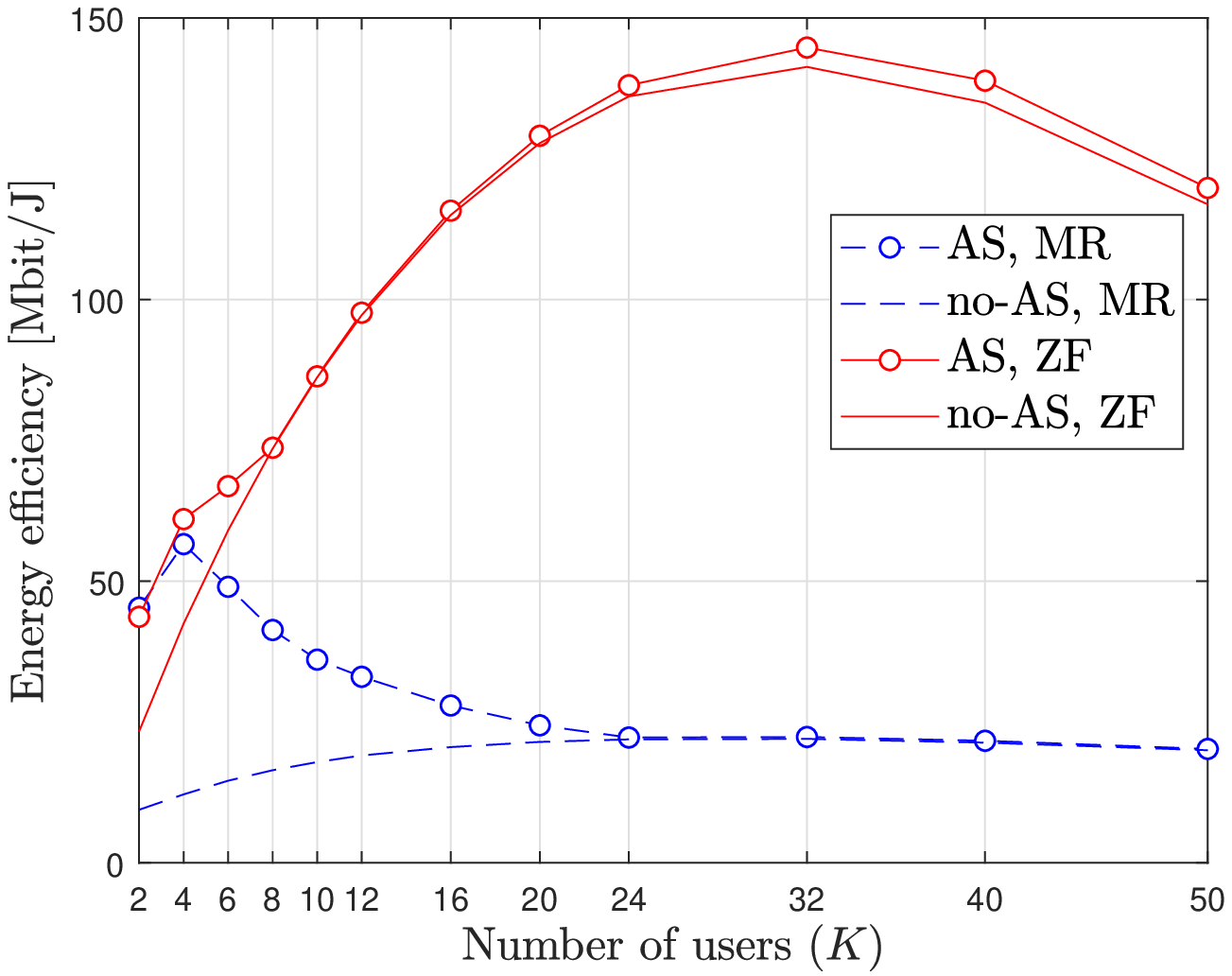}
		\caption{Energy efficiency}
		\label{fig:EExK}
	\end{subfigure}%
	\begin{subfigure}{.5\textwidth}
		\centering
		\includegraphics[width=\textwidth]{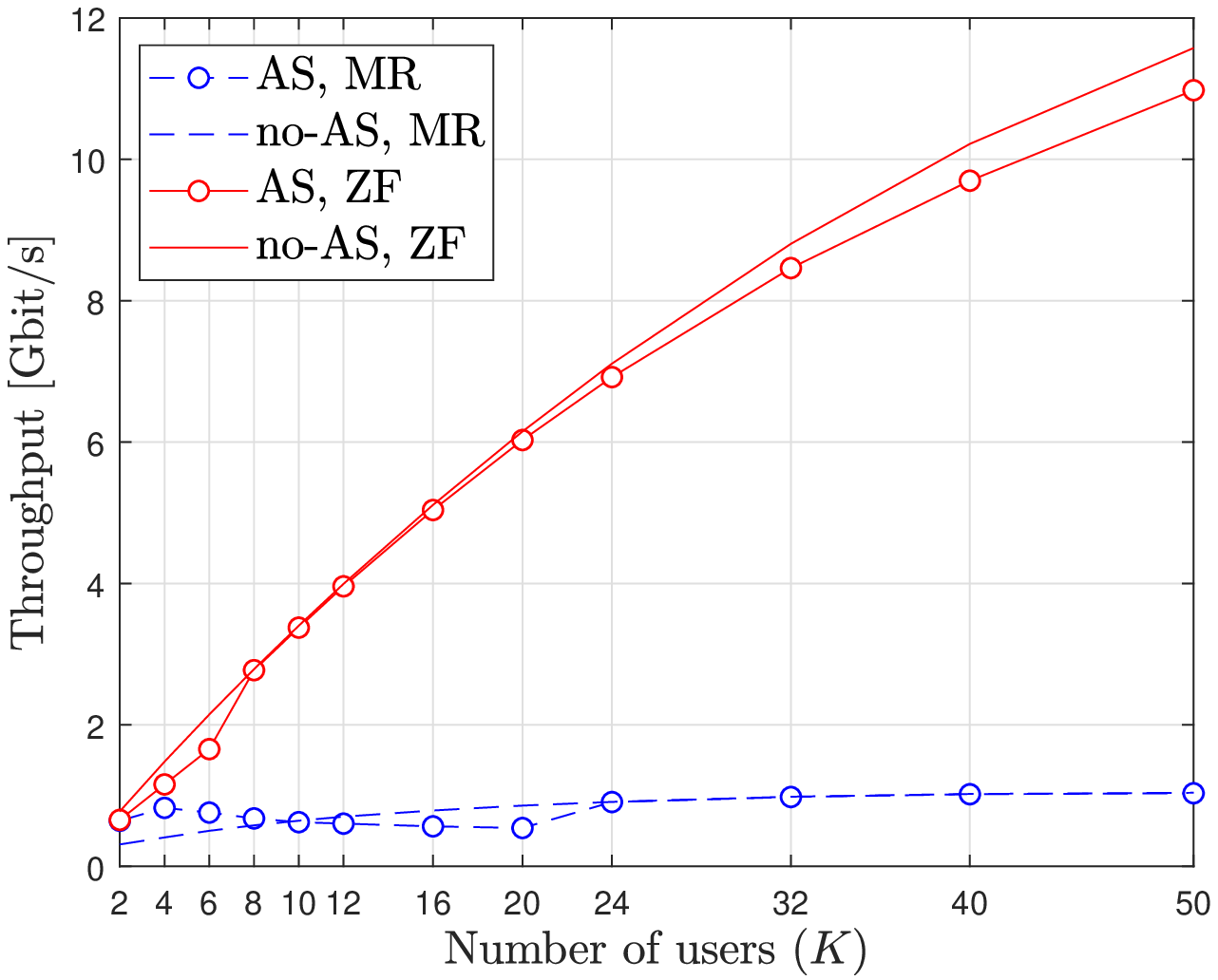}
		\caption{Throughput (sum rate)}
		\label{fig:THROUGHPUTxK}
	\end{subfigure}
	\newline
	\begin{subfigure}{.5\textwidth}
		\centering
		\includegraphics[width=\textwidth]{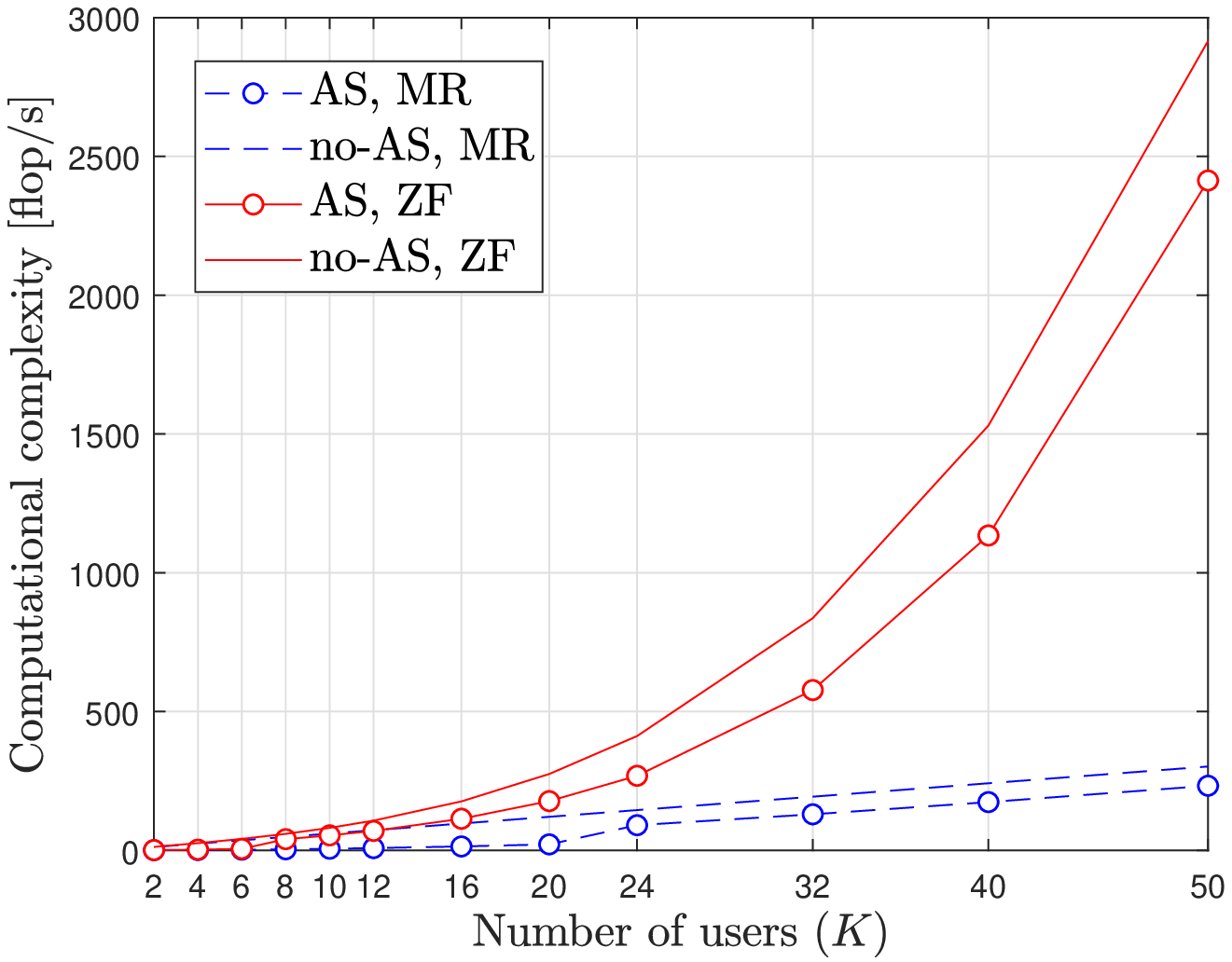}
		\caption{Total computational complexity}
		\label{fig:CxK}
	\end{subfigure}%
	\begin{subfigure}{.5\textwidth}
		\centering
		\includegraphics[width=\textwidth]{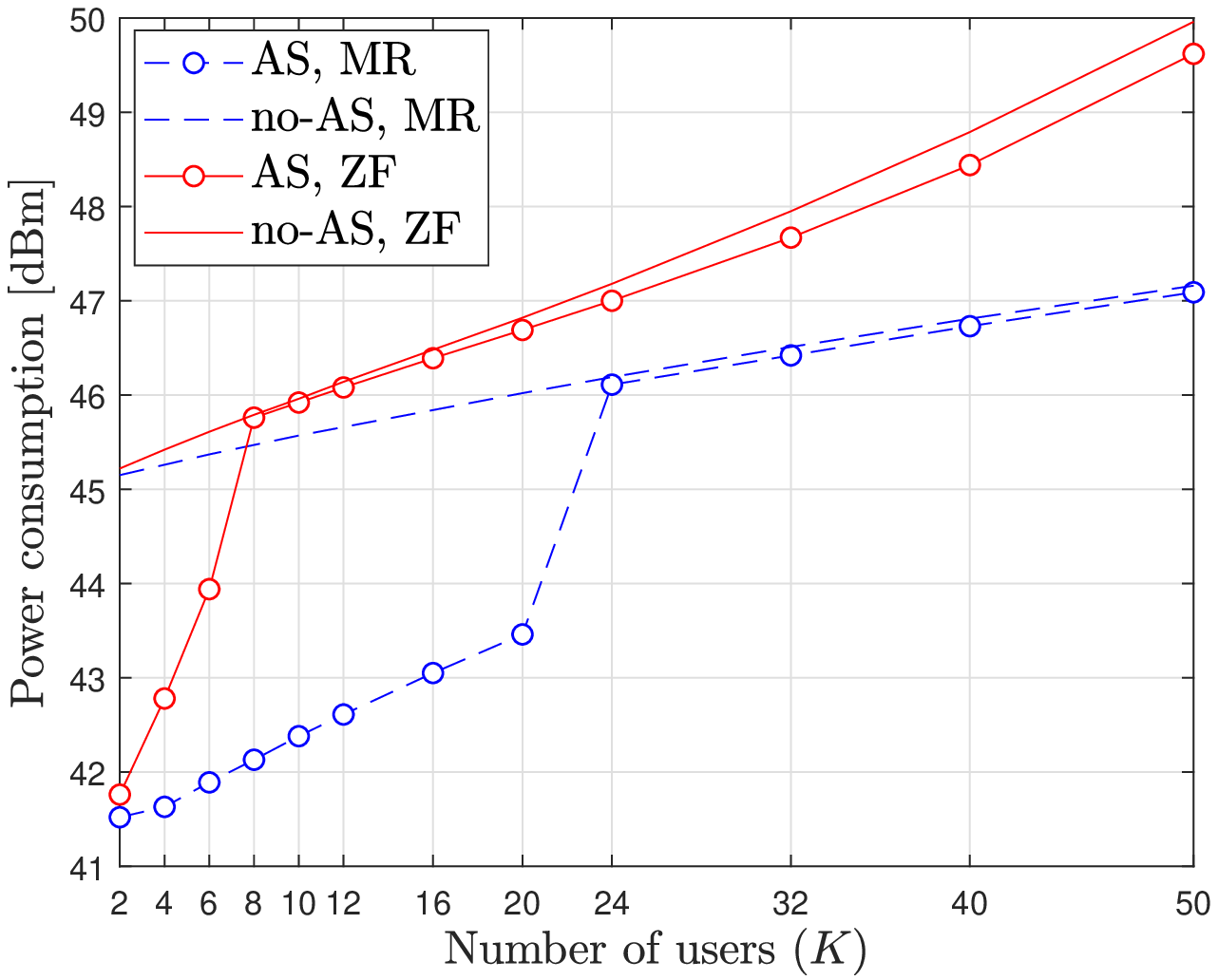}
		\caption{Total power consumption}
		\label{fig:PxK}
	\end{subfigure}%
	\caption{{EE, throughput, power consumption and computational complexity for the HRNP-AS and and  no-AS strategies taking $N=N^*$.}}
	\label{fig:xK}
	\end{figure*}

{Figs. \ref{fig:EExK}, \ref{fig:THROUGHPUTxK}, \ref{fig:CxK} and \ref{fig:PxK} depict the EE, throughput, power consumption and computational complexity, respectively, for the proposed HRNP-AS and no-AS strategy assuming $N=N^*$. If the system operates under the bound $K\leq K_\text{max}$, the HRNP-AS strategy provides an increasing in the EE and simultaneously reduces the power consumption and the computational burden, while the throughput is very close to the obtained with the no-AS strategy.}

{Besides, when increasing $K$ until it is close to $K_\text{max}$, the EE gradually decreases until it meets the no-AS strategy EE curve. It occurs because taking $N \leq M/10$ is no longer enough to mitigate the interference. Besides, from the point $K=K_\text{max}$, $N^*$ suddenly jumps to about 50 or 60\% of the ULA array size, $M=100$. Finally, when the XL-MIMO system operates over the maximum number of user bound, $K>K_\text{max}$, the AS-strategy is no longer able to improve the EE and simultaneously reduce the power consumption considerably. However, it is still advantageous when compared to the all-antennas activation strategy, as it is still able to reduce the computational burden.}

\section{Conclusion}\label{sec:conclusion}
	%
{In XL-MIMO systems operating under non-stationary channels, the users see only a portion of the antenna array and the majority of the energy sent by the users is concentrated on this part of the array. Therefore, by appropriately selecting a subset of antennas that communicate to each user, it can be guaranteed to capture almost the totality of the incident energy from that user, while reducing substantially the interference coming from the other $(K-1)$ users. As corroborated by extensive numerical results, the HRNP antenna selection criterion reduces considerably the complexity of computing the combiners and precoders and consequently reducing the power consumption also. Furthermore, when the AS-HRNP algorithm is set to select only a few $N$ antennas per user, there may be several antennas which are not activated, reducing the transceiver chains power consumption. As the HRNP-based AS strategy results in almost the same spectral efficiency as the no-AS strategy and a considerable substantial power consumption reduction, as a result, the EE is increased significantly. Furthermore, the extensive numerical results demonstrated the existence of an optimal value of $N$ in terms of maximizing the EE, which depends on the number of users and array size. Also, it is not even advantageous to increase $N$ beyond this optimal value, since neither the throughput nor the EE would be considerably improved while computational burden and energy consumption increases remarkably.}



\end{document}